\documentclass[fleqn,usenatbib]{mnras}

\usepackage{newtxtext,newtxmath}
\usepackage{comment}

\usepackage[T1]{fontenc}

\DeclareRobustCommand{\VAN}[3]{#2}
\let\VANthebibliography\thebibliography
\def\thebibliography{\DeclareRobustCommand{\VAN}[3]{##3}\VANthebibliography}

\usepackage{graphicx}	
\usepackage{amsmath}	
\usepackage{tabularx}
\newcommand{\gray}{$\gamma$-ray}
\newcommand{\grays}{$\gamma$-rays}

\title[BL Lacs and FSRQs in 4FGL]{Gradient boosting decision trees classification of blazars of uncertain type in the fourth Fermi-LAT catalog}

\author[N. Sahakyan et al.]{
N. Sahakyan,$^{1,2}$ \thanks{E-mail: narek@icra.it}
V. Vardanyan,$^{1}$
M. Khachatryan $^{1}$
\\
$^{1}$ICRANet-Armenia, Marshall Baghramian Avenue 24a, Yerevan 0019, Armenia\\
$^{2}$ICRANet, P.zza della Repubblica 10, 65122 Pescara, Italy\\
}
\date{Accepted XXX. Received YYY; in original form ZZZ}

\pubyear{2015}

\begin{document}
\label{firstpage}
\pagerange{\pageref{firstpage}--\pageref{lastpage}}
                        \maketitle

\begin{abstract}
The deepest all-sky survey available in the $\gamma$-ray band - the last release of the Fermi-LAT catalogue (4FGL-DR3) based on the data accumulated in 12 years, contains more than 6600 sources. The largest population among the sources is blazar subclass - 3743, $60.1\%$ of which are classified as BL Lacertae objects (BL Lacs) or Flat Spectrum Radio Quasars (FSRQs), while the rest are listed as blazar candidates of uncertain type (BCU) as their firm optical classification is lacking. The goal of this study is to classify BCUs using different machine learning algorithms which are trained on the spectral and temporal properties of already classified BL Lacs and FSRQs. Artificial Neural Networks, \textit{XGBoost} and \textit{LightGBM} algorithms are employed to construct predictive models for BCU classification. Using 18 input parameters of 2219 BL Lacs and FSRQs, we train (80\% of the sample) and test (20\%) these algorithms and find that \textit{LightGBM} model, state-of-the-art classification algorithm based on gradient boosting decision trees, provides the highest performance. Based on our best model, we classify 825 BCUs as BL Lac candidates and 405 as FSRQ candidates, however, 190 remain without a clear prediction but the percentage of BCUs in 4FGL is reduced to 5.1\%. The $\gamma$-ray photon index, synchrotron peak frequency, and high energy peak frequency of a large sample are used to investigate the relationship between  FSRQs and BL Lacs (LBLs, IBLs, and HBLs). 
\end{abstract}




\section{Introduction} 
Fermi Gamma Ray Space Telescope (Fermi) launched in 2008 with improved sensitivity, wide field of view, large energy range, and an all-sky-survey operation mode, provided an unprecedented detailed view of the \gray\ sky. The primary instrument onboard the spacecraft, the Large Area Telescope (LAT), performs all sky survey every three hours in the energy range from $\sim100$ MeV to $>300$ GeV providing continuous and deepest view of the \gray\ sky.  Further details on Fermi-LAT are given in \citet{2009ApJ...697.1071A}. 

The Fermi-LAT observations resulted in detection of many galactic and extragalactic \gray\ sources. For example, the most recent fourth Fermi-LAT catalog of \gray\ sources 
\citep[4FGL Data Release 3 (DR 3)][]{2022ApJS..260...53A}
based on the data accumulated between 2008-2020 (12 years) contains 6659 Galactic and extragalactic source of different classes. Pulsars are the largest class of Galactic \gray\ emitters - 292, and other Galactic \gray\ emitters are: globular clusters- 25, Supernova remnants - 43 and Pulsar wind nebulae - 19. The extragalactic \gray\ sky is largely dominated by active galactic nuclei, in particular by blazars which also represent the largest fraction of the sources in 4FGL, 3743 out of 6659.

Blazars are a rare type of AGNs when one of the jets makes a small angle $(<10^\circ)$ to the line of sight of the observer \citep{1995PASP..107..803U}. The nonthermal emission from blazars is characterized by rapid flux variability across the entire accessible electromagnetic spectrum, the most extreme being at \grays. Blazars are sub-grouped in two large classes, Flat Spectrum Radio Quasars (FSRQs) and BL Lacertae objects (BL Lacs), based on the properties observed in the optical spectrum, namely, in FSRQs strong and quasar-like emission lines are observed, whereas in BL Lacs the emission lines are weak or absent \citep{1995PASP..107..803U}. Blazars are further classified using the observed frequency of the synchrotron peak ($\nu_{\rm s,p}$) as either low synchrotron peaked sources (LSPs or LBLs) when $\nu_{\rm s,p}<10^{14}$ Hz, intermediate synchrotron peaked sources (ISPs or IBLs) when $10^{14}$ Hz $<\nu_{\rm s,p}<10^{15}$ Hz and high synchrotron peaked sources (HSPs or HBLs) when $\nu_{\rm s,p}>10^{15}$ Hz. In this classification FSRQs have $\nu_{\rm s,p}$ similar to those of LBLs. Among the blazars included in 4FGL, 1456 are BL Lacs, 794 are FSRQs and 1493 are blazar candidates of uncertain type (BCU). BCUs display properties similar to blazars (e.g., a flat radio spectrum and a typical two-humped blazar-like spectral energy distribution), but reliable optical association is lacking.

BCUs corresponds to the 39.9\% of all blazars included in the 4FGL and their possible classification is very important for the scientific community, as it can be useful for blazar population studies (i.e., properties of different blazar sub-classes) or for planning observational campaigns on individual objects. Moreover, possible classification of BCUs is also important, considering the recent possible association of blazars and IceCube events that triggered interest on possible multi-messenger observations of blazars \citep[e.g.,][]{2018Sci...361.1378I, 2018Sci...361..147I,2018MNRAS.480..192P, 2022arXiv220405060S}. 
Although various optical monitorings aim to classify BCUs, they are time consuming and costly, given the large number of BCUs. However, in recent years there is a growing interest in applying machine learning techniques to different fields of science, including astronomy and astrophysics. Machine learning is a powerful tool in data science allowing machines to learn from data, detect patterns, self-improve and make classifications. The distinct spectral properties of BL Lacs and FSRQs in the \gray\ band can be used to train models which then can classify BCUs by comparing their properties with those of BL Lacs and FSRQs. 

In fact, machine learning was already applied by \citet{2012ApJ...753...83A, 2016MNRAS.462.3180C, 2016ApJ...820....8S, 2017MNRAS.470.1291S, 2017A&A...602A..86L, 2019MNRAS.490.4770K, 2020MNRAS.498.1750A, 2020MNRAS.493.1926K, 2020ApJ...895..133X, 2021MNRAS.507.4061F, 2021MNRAS.505.5853G, 2021JHEAp..29...40C, 2021RAA....21...15Z, 2021MNRAS.505.1268F, 2022ApJS..259...55N, 2022A&A...660A..87B, 2022JCAP...04..023B, 2022A&A...660A..87B, 2022MNRAS.515.1807C} and other authors to study the multiwavelength properties of blazars, or classify unassociated \gray\ sources, or classify BL Lacs and FSRQs among BCUs. The methods used to identify or classify blazars include Artificial Neural Network (ANN) \citep[e.g.,][]{2016MNRAS.462.3180C, 2017MNRAS.470.1291S, 2019MNRAS.490.4770K, 2020MNRAS.493.1926K}, multivariate classifiers - boosted decision trees and multilayer perceptron neural network \citep{2017A&A...602A..86L}, Bayesian Neural Networks \citep{2022JCAP...04..023B}, Random Forest \citep[e.g.,][]{2016ApJ...820....8S}, CatBoost gradient boosting decision trees \citep{2022MNRAS.515.1807C}, and others. All the applied models with different performances produce satisfactory results in classifying blazars based on different properties. It should be noted that machine learning techniques are also used to estimate different properties of blazars. For example, in \citet{2016MNRAS.462.3180C} ANN was used to identify high synchrotron peaked blazars or \citet{2022arXiv220703813G} developed BlaST which uses machine learning methods to estimate the synchrotron peak directly from the blazar spectral energy distribution.

\begin{figure*}
    \centering
    \includegraphics[width=0.98\textwidth]{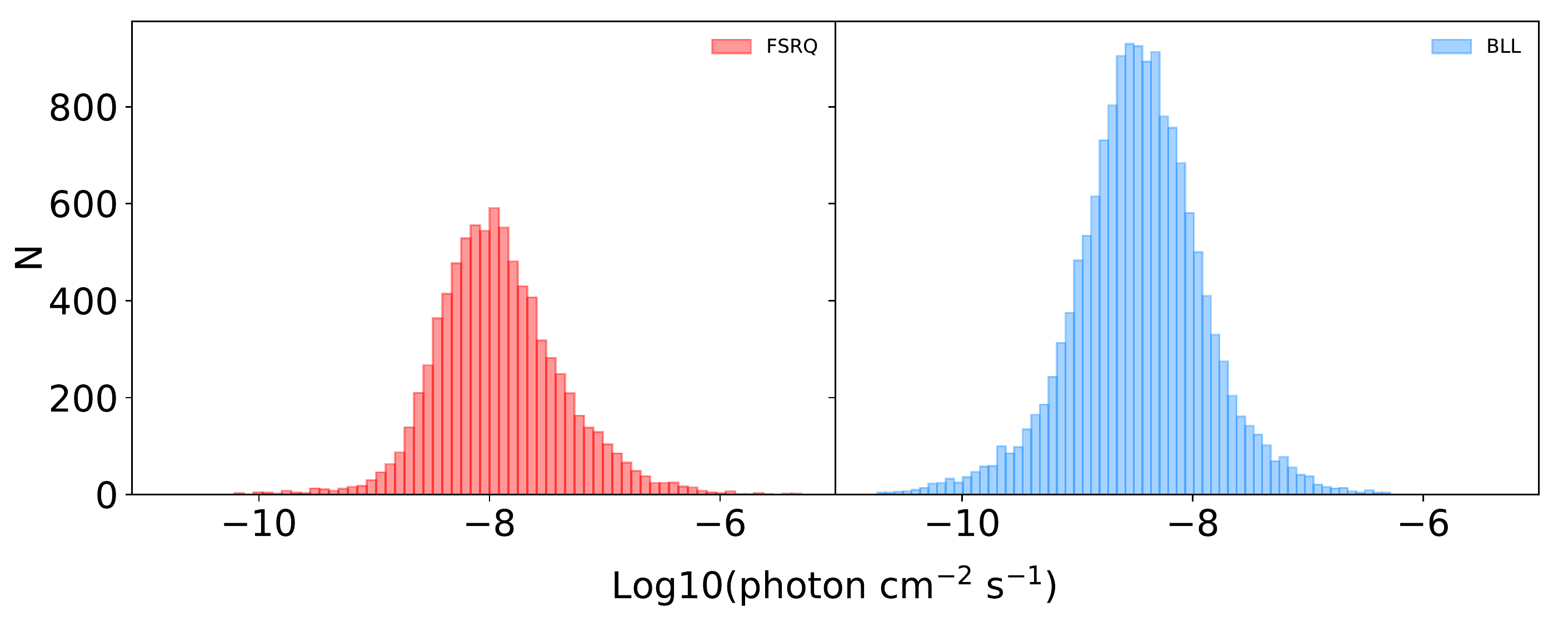}
    \includegraphics[width=0.98\textwidth]{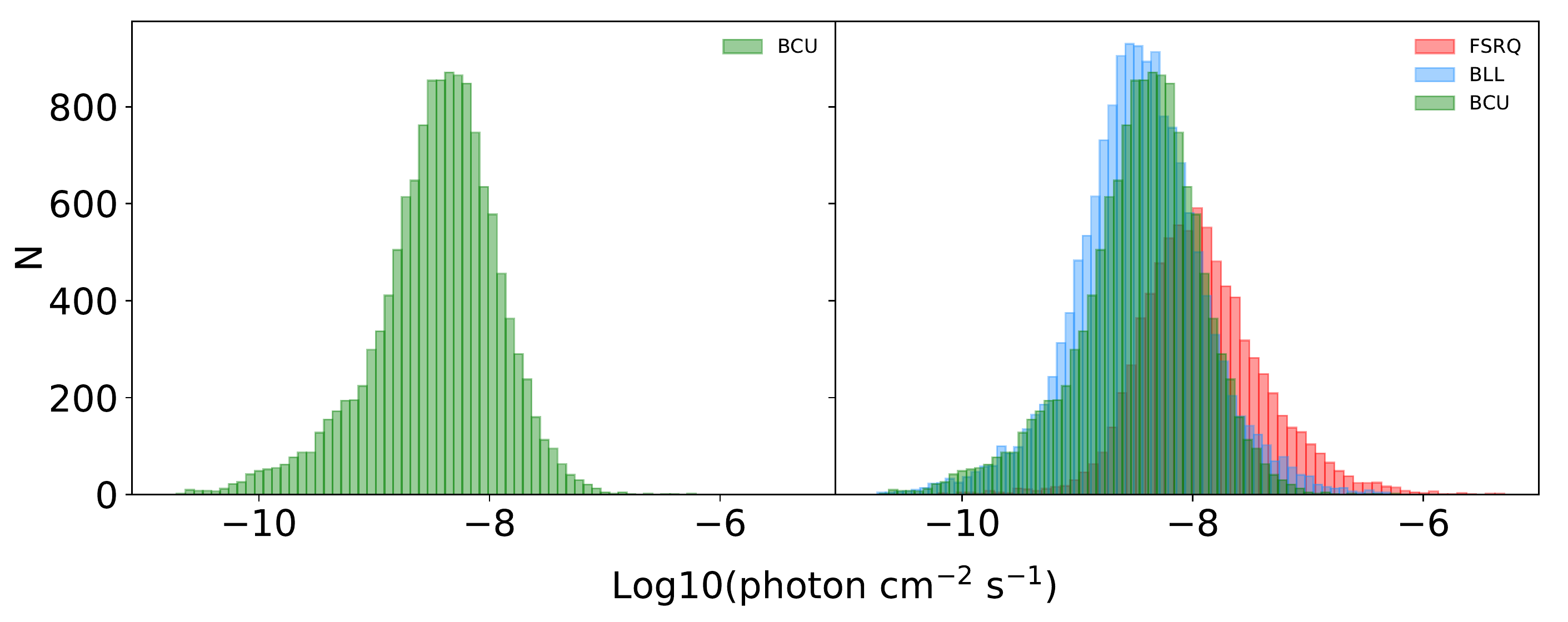}
    \includegraphics[width=0.98\textwidth]{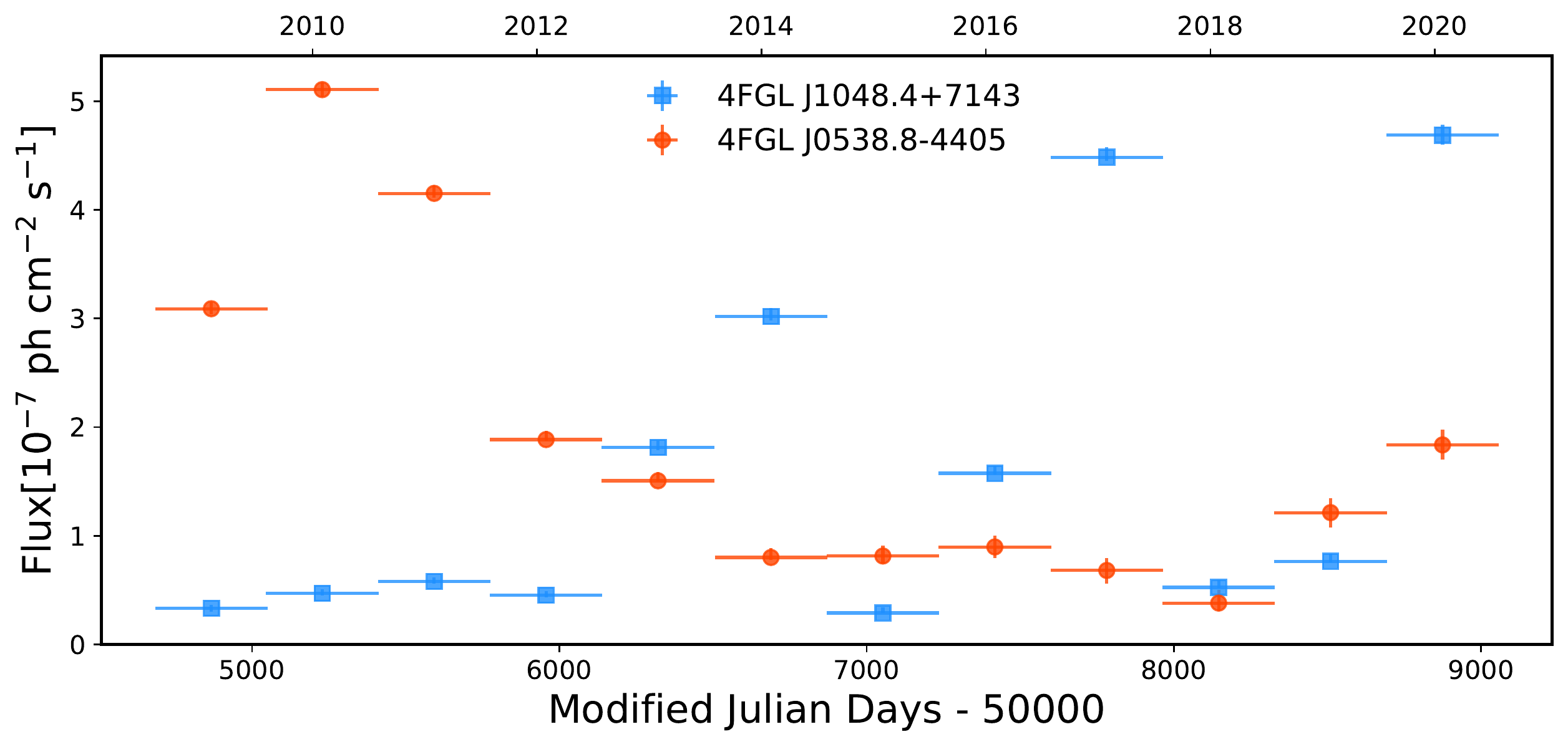}
    \caption{\textit{Upper and middle panels:} The distribution of yearly fluxes of FSRQs, BL Lacs and BCUs included in the 4FGL catalog. \textit{Lower panel:} The light curves of 4FGL J0538.8-4405 (BL Lac) and 4FGL J1048.4+7143 (FSRQ) measured during 12 years (2008-2020).}
    \label{blazdistri}
\end{figure*}
The goal of the current work is to perform a machine learning classification of BCUs by training the models on the most up-to-date \gray\ dataset which contains spectral and temporal properties of BL Lacs and FSRQs. Using the most complete \gray\ dataset based on 12 years of Fermi-LAT operation (4FGL) as well as state-of-the-art classification methods (eXtreme Gradient Boost \citep{2016arXiv160302754C} and \textit{LightGBM} \citep{NIPS2017_6449f44a}) will better identify common patterns in \gray\ properties of BL Lacs and FSRQs and will allow statistically better association of BCUs.

The structure of the paper is as follows. In Section \ref{sources}, the source sample and the properties of the used data are presented. The various algorithms used in the current paper are discussed in Section \ref{model}. In Section \ref{res}, we provide an overview of the obtained results and provide the conclusions in Section \ref{conc}.
\section{The source sample from 4FGL-DR3}\label{sources}
The incremental version of the fourth Fermi-LAT catalog (4FGL Data Release 3 [DR3]) contains 6659 sources and for each individual source together with the coordinates, various spectral properties are provided, such as flux, detection significance, spectral parameters when fitting with different models, etc. For our study, from 4FGL DR3 we selected  all sources with BLL, FSRQ and bll, fsrq designations where in capital letters are firm identifications whereas lower case letters indicate associations \citep[see][]{2022ApJS..260...53A}. This amounted to 2250 sources (1456 BL Lacs and 794 FSRQs) to train our models.
Then those models are applied to classify 1493 blazar candidates.\\
\indent Among the features and measurements presented in 4FGL, we are interested in the energy spectra and fluxes measured in different periods; the first  correspond to the sources' fluxes measured in different energy bands (\textit{nuFnu\_band} column) while the second one represent the sources' flux as a function of time (\textit{Flux\_History} column). These two measurements well characterize different blazar sub-classes and are used as input parameters for our models.    
\subsection{$\gamma$-ray light curves}
4FGL contains information on the \gray\ flux of the considered sources in different time bins (light curve) which is an essential information on the blazars emission features observed in the initial 12 years of Fermi-LAT operation. These light curves were computed by dividing the whole time period into twelve intervals (one year each) and the fluxes in each sub-interval were estimated by applying binned (up to 10 GeV) and unbinned (10-100 GeV) likelihood analysis. These fluxes were estimated by freezing the spectral parameters to those obtained in the fit over the full range and by adjusting the normalization. The photon fluxes estimated in the energy range 0.1-100 GeV in the units of ${\rm photon\:cm^{-2}\:s^{-1}}$ are reported for each year.

For each selected source this creates twelve parameters that describe the \gray\ flux evolution in different periods. One may speculate that the short time flux variability will be smoothed out when measuring the flux in one-year intervals. However, the goal is to identify common patterns in the change of the \gray\ flux of blazars in different sub-classes rather than to compare the variability timescales. Moreover, by comparing the variability of blazars using 2-month and 1 year light curves, \citet{2020ApJS..247...33A} showed that among 1173 sources identified as variable in 2-month intervals, 1057 show variability also in yearly binned intervals. Therefore, the one-year-binned light curves contain most of the variability information on the considered sources. Also, when the flux is measured in shorter periods, it might result in many upper limits which creates additional uncertainties for the models.

The distribution of the yearly measured fluxes of BL Lacs, FSRQs and BCUs is shown in Fig \ref{blazdistri} (upper and middle panels), highlighting the difference in their \gray\ emission. For example, the mean of FSRQ \gray\ fluxes distribution is at $3.28 \times 10^{-8}\:{\rm photon\:cm^{-2}\:s^{-1}}$ while that of BL Lacs is at $8.19\times 10^{-9}\:{\rm photon\:cm^{-2}\:s^{-1}}$. FSRQs are brighter with a highest yearly flux of $5.05 \times 10^{-6}\:{\rm photon\:cm^{-2}\:s^{-1}}$ observed for 4FGL J2253.9+1609 (3C 454.3) as compared with the similar value of $5.2 \times 10^{-7}\:{\rm photon\:cm^{-2}\:s^{-1}}$ observed for BL Lacs, 4FGL J2202.7+4216 (BL Lacertae). The BCU yearly fluxes distribution with a mean of $6.72\times 10^{-9}\:{\rm photon\:cm^{-2}\:s^{-1}}$ is broader, mimicking the properties of both FSRQs and BL Lacs. The distribution of yearly fluxes of FSRQs, BL Lacs and BCUs are shown together in Fig. \ref{blazdistri} middle panel right side.

The fluxes estimated in each interval represent an independent state of the sources and show their brightest and lowest emission states. The light curves of the sources (FSRQ and BL Lac) shown in Fig. \ref{blazdistri} (lower panel) clearly demonstrate their different emission states. For example, BL Lac 4FGL J0538.8-4405 (orange circles in Fig. \ref{blazdistri} lower panel) is initially in the high \gray\ emission state while it is in a relatively faint state in 2014, but then brightens again at the end of the considered period. Instead, the emission from FSRQ 4FGL J1048.4+7143 (blue squares in Fig. \ref{blazdistri} lower panel) is initially in a relatively faint state but then it is in repeatedly flaring and quiescent states.
Therefore, given that  blazar emission is variable, a simple comparison of the fluxes of different sources estimated in the same year does not have any physical motivation. On the other hand, the change in the flux is linked with the source properties and it is meaningful to compare the fluxes when the sources are in the lowest, average or brightest emission states. Thus, we have sorted the early measured fluxes from the lowest to the highest and they are considered as twelve different input parameters. So, the network can compare and contrast the fluxes of sources whether they are in low or bright emission states. Some sources do not have the flux measured for all yearly bins, only the upper limits are given, so these data are missing. For the periods with no detection (upper limit) zero or NaN were set, depending on the method, meaning that there is no information.
\begin{figure*}
    \centering
    \includegraphics[width=0.49\textwidth]{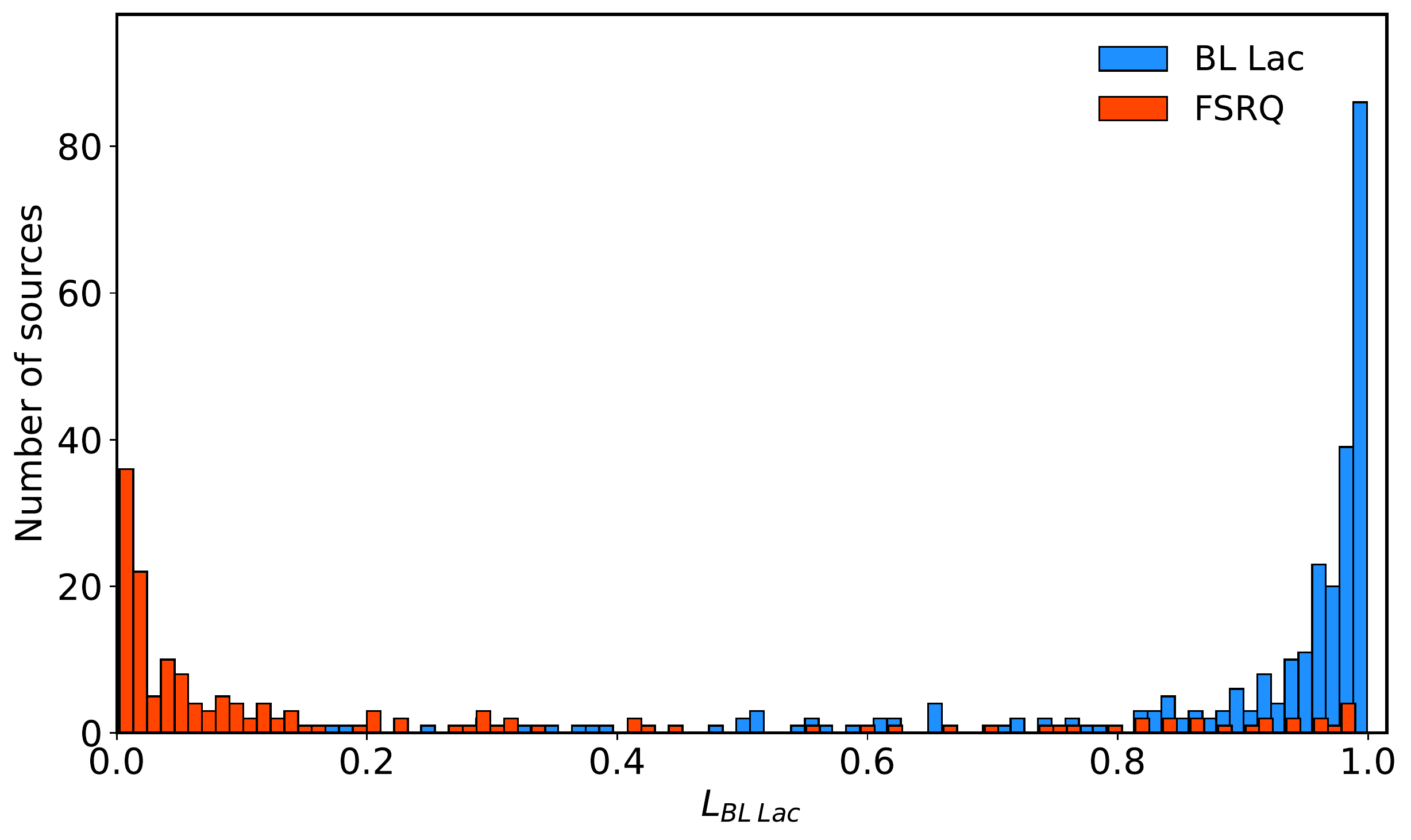}
    \includegraphics[width=0.49\textwidth]{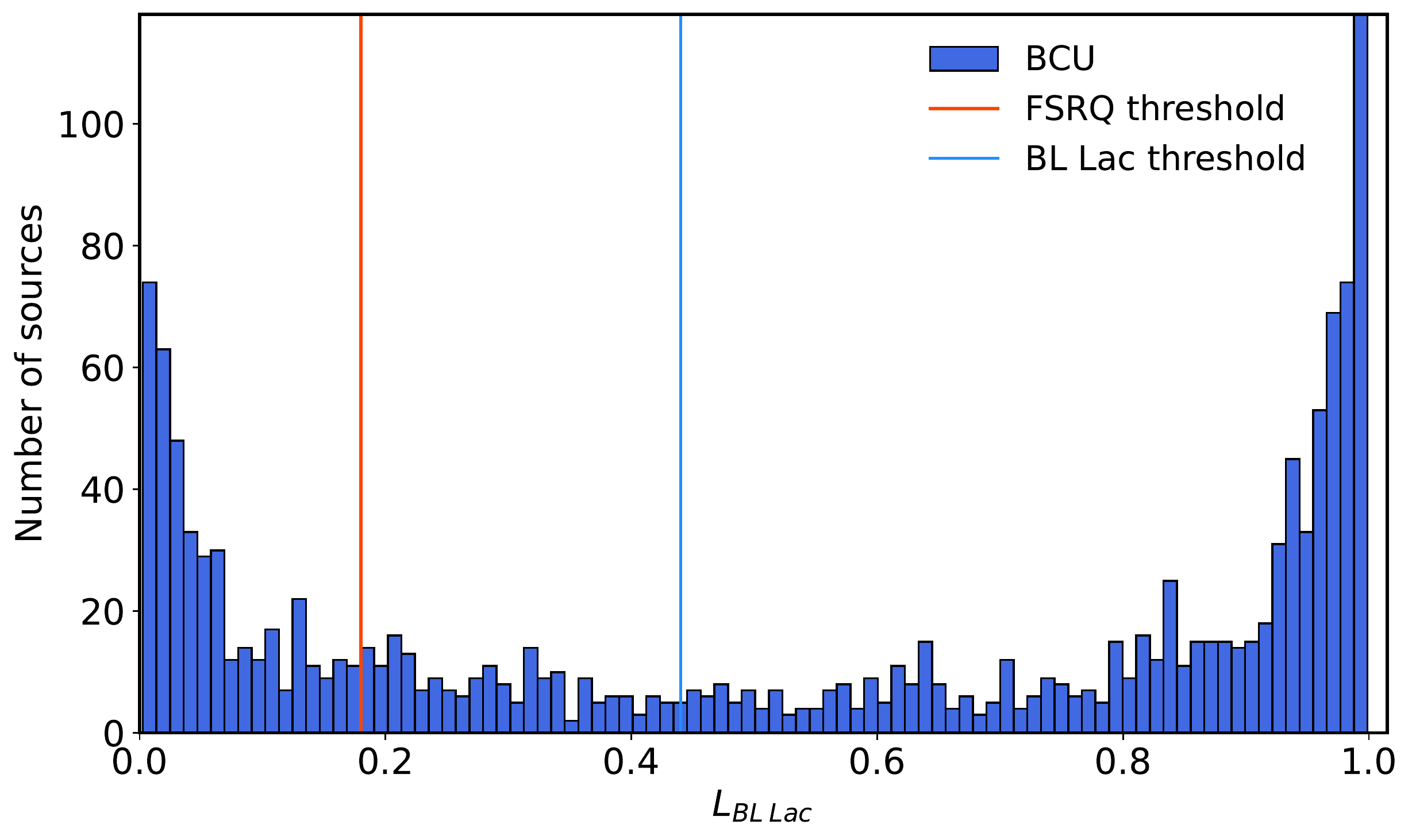}
    \caption{\textit{Left panel:} Distribution of the likelihood to be BL Lac (blue) or FSRQ (orange) for the sources in the test sample. \textit{Right panel:} Distribution of the likelihood of 1420 BCUs to be BL Lac or FSRQ candidates.}
    \label{likelihood}
\end{figure*}
\subsection{$\gamma$-ray spectra}
It is known that the \gray\ spectra of BL Lacs and FSRQs shown distinct differences. For example, \citet{2020ApJS..247...33A} have shown that 93\% of FSRQs have power-law photon indices $>2.2$ while those of 81\% of BL Lacs $<2.2$. However, the photon index of BL Lacs varies also for LBLs, IBLs and HBLs; the \gray\ spectra of LBLs are softer than those of HBLs \citep[see e.g.,][]{2020ApJ...892..105A}. In principle, the difference between FSRQs and BL Lacs can be of a purely physical origin. In BL Lacs jets the electrons can be accelerated to higher energies, having a harder energy spectrum and hence producing photons with harder spectra. Instead, in FSRQs where the electrons effectively interact with different photon fields and efficiently cool down, the produced photons will appear with a soft \gray\ spectrum. Therefore, the spectral difference between FSRQs and BL Lacs can be used for BCU classification. 

The 4FGL catalogue provides the sources fluxes (spectra) measured in eight energy bands: \textit{1)} 50–100 MeV, \textit{2)} 100–300 MeV, \textit{3)} 300 MeV–1 GeV, \textit{4)} 1–3 GeV, \textit{5)} 3–10 GeV, \textit{6)} 10–30 GeV, \textit{7)} 30–100 GeV and \textit{8)} 100 GeV–1 TeV. We excluded the first band, as at lower energies the fluxes could be affected by contamination of other sources.
Visually inspecting plots of spectral fits of all sources, we decided to drop the last band as well because it is an upper limit for many sources, and work with the remaining 6 bands. The fluxes in each of these bands contain information of average spectra of the sources (e.g., spectral index, spectral curvature, spectral breaks, etc.) and can be used to distinguish between different types of blazars. The fluxes measured in each energy band were not sorted and provided as input according to the increase of the energy because they are defined by the photon index which is different for FSRQs and BL Lacs. There is already a physical interpretation for the fluxes in the same input parameter, so their compression is meaningful.
\begin{figure}
    \centering
    \includegraphics[width=0.45\textwidth]{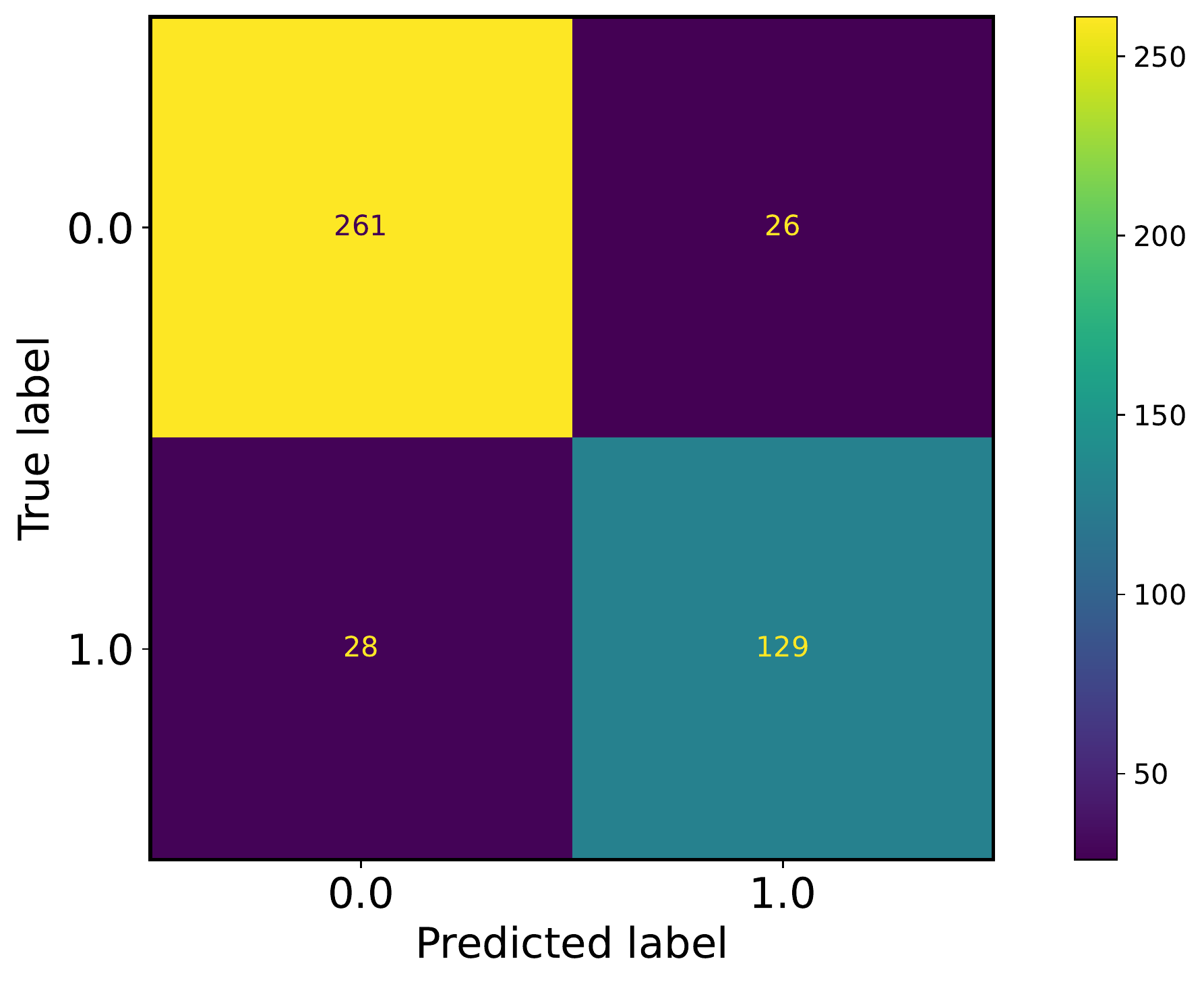}
    \caption{The confusion matrix of \textit{LightGBM\_opt} classifier on the test sample.}
    \label{conf_matrix}
\end{figure}
\section{Model construction}\label{model}
The aim of this work is to examine the nature of BCUs in 4FGL based on their \gray\ properties. The spectral and temporal properties discussed in the previous section provide a framework for predicting the expected types of unclassified blazars. This is done by defining models that find correlations between measured \gray\ properties of BL Lacs and FSRQs and then compare them to the \gray\ properties of BCUs. Here, we have implemented two different machine learning techniques to classify BCUs: Artificial Neural Network (ANN) and Gradient Boosted Decision Tree algorithm. Below we briefly introduce the general features of the used techniques.

\begin{figure*}
    \centering
    \includegraphics[width=0.49\textwidth]{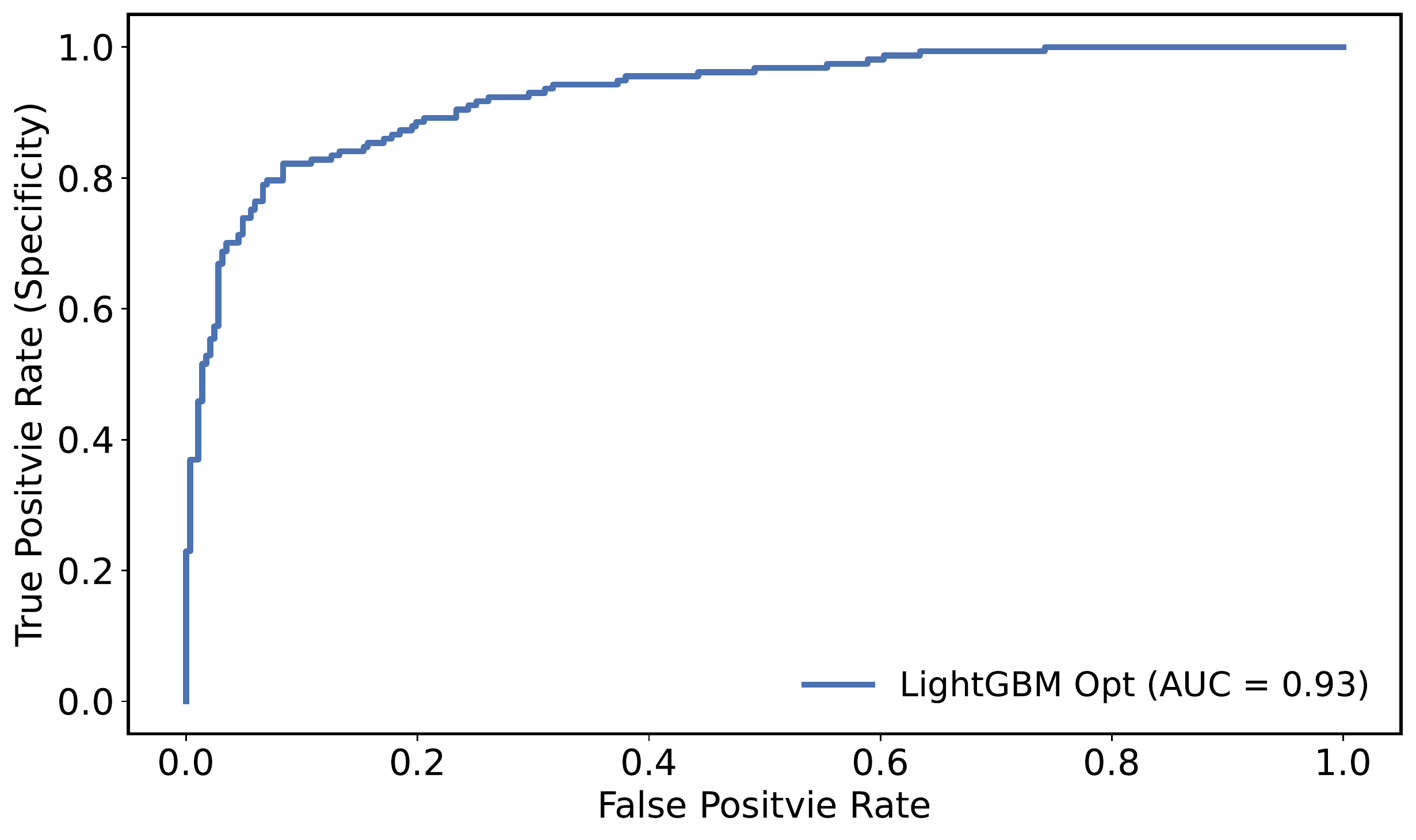}
    \includegraphics[width=0.49\textwidth]{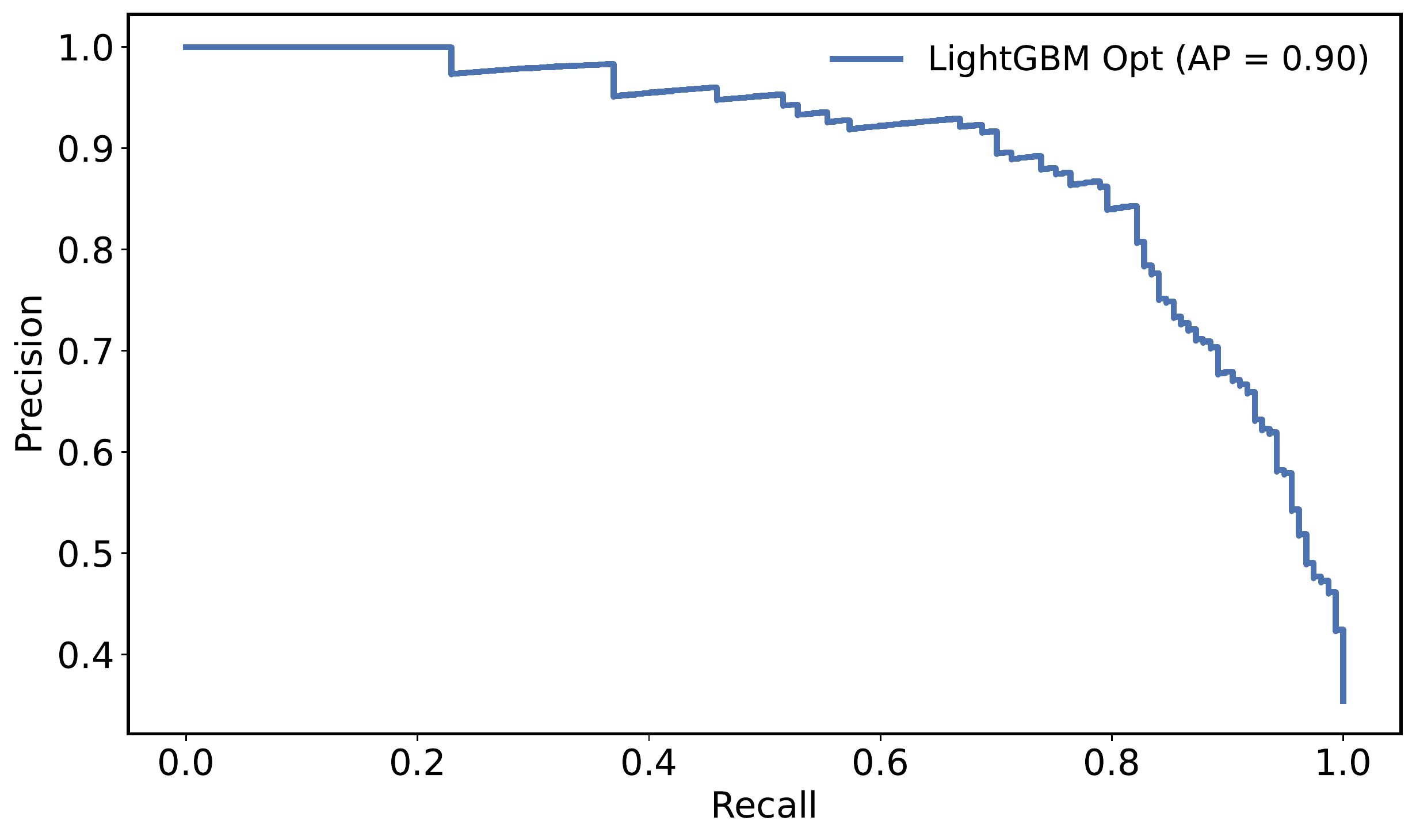}
    \caption{\textit{Left panel:} ROC curve for \textit{LightGBM\_opt} classifier. \textit{Right panel:} Precision Recall curve. }
    \label{auc}
\end{figure*}
\begin{table*}
\begin{center}
\caption{The BCU classification performance of the models.}
\begin{tabular}{ c|c|c|c|c } 
 \hline \hline
 Model & Recall of Minority & Recall Weighted & Precision Weighted & ROC-AUC Macro \\
 \hline
 ANN &   0.81 & 0.88 & 0.88 & 0.86 \\ 
 \textit{XGBoost\_def} & 0.77 & 0.87 & 0.87 & 0.85 \\
 \textit{XGBoost\_opt} & 0.80 & 0.87 & 0.87 & 0.86 \\

 \textit{LightGBM\_def} & 0.81 & 0.87 & 0.87 & 0.86 \\
 \textit{LightGBM\_opt} & 0.82 & 0.88 & 0.88 & 0.87 \\
 \hline \hline

\end{tabular}
\label{aaa}
\end{center}

\end{table*}
\begin{itemize}
  \item \textit{Artificial Neural Network}
  
  Artificial Neural Networks (ANNs) \citep{bishop1995neural} are among the most powerful tools in pattern-recognition problems. ANNs have been used successfully in various fields, including astrophysics and cosmology \citep[e.g.,][]{10.1111/j.1745-3933.2006.00276.x, 10.1093/mnras/stv632, pnas.1821458116, 2017MNRAS.470.1291S, 2016MNRAS.462.3180C, 2019MNRAS.490.4770K,  2020MNRAS.493.1926K}. 
  
  ANN consists of input, hidden, and output layers with connected neurons (nodes) representing a simplified model of the human brain functioning and the nervous system. A standard neural network contains an input layer and an output layer but in most cases it can include any number of hidden layers with any number of hidden nodes in each layer. The input parameters, neurons in the input layer, are connected to one or more neurons in the hidden layer (intermediate layer) and propagate the data to the deeper layers and send the final output data to the last output layer (prediction). Each neuron in the first hidden layer is assigned weights associated to input parameters which indicates the importance of each neuron in the network. The goal of ANN training is to minimize the output error by finding the best set of weights for each connection. Initially, the weights are assigned randomly and are optimized during an ANN training. So, the ANN uses the input data to produce an output data which is compared with the real data to calculate the error (loss function). Then, ANN learns by adjusting its weights such that in the next iteration the net error produced by the ANN is generally smaller than that in the current iteration. So, it optimizes the weight values to get the best result from the network.
  
  In our case, when the classification is the goal, the input parameters (fluxes in different years and fluxes in each energy bin) are values describing blazars while output layer is the number of classes (FSRQ or BL Lac). The network is trained (i.e., to find a function which best separates objects belonging to different classes) on already classified BL Lacs and FSRQs, tested on a selected sample of classified sources that was not used in the training, and then the resultant model can be used to classify BCUs.
  \item \textit{Gradient Boosted Decision Tree}
  
  Gradient Boosted Decision Tree (GBDT) is a machine learning algorithm used for both classification and regression problems. Boosting is one technique which aims to build a strong classifier from a number of weak classifiers, so it is forward-learning ensemble method that obtains results by gradually improving the estimations.  Initially, a model (e.g., a tree) is fitted to the data, and then a second model is constructed by improving the cases where the accuracy of the first model was not good. Then, these processes of boosting are repeated many times to create a series of decision trees that produce an ensemble of weak prediction models; each successive model attempts to correct for the weakness of all the previous models and the combination of new models is better than the previous ones alone. In the gradient boosting method the loss function is minimized by adding trees in a gradient descent procedure. Namely, the very first model is trained on the dataset, whereas the second model is trained on the errors of the first model and added to the first model and so on. GBDT algorithms have wide applications and been used also in astronomy and astrophysics in a variety of problems \citep[e.g.,][]{2022MNRAS.515.1807C, 2021MNRAS.503.4136G, 2019MNRAS.485.4539J, 2019ApJ...887..241Y}. 
  
  Here we use \textit{XGBoost} and \textit{LightGBM} methods based on gradient boosting algorithm to classify BCUs. Extreme Gradient Boosting (XGBoost) \citep{2016arXiv160302754C} is a scalable machine learning algorithm for tree boosting where the best model is found by applying more accurate approximations. Unlike GBDT, in \textit{XGBoost} model the objective function is optimized using Newton-Raphson method, i.e., second-order partial derivatives are used to gather more information about the direction of the gradient and the way to get to the minimum of the loss function. \textit{LightGBM }is another implementation of GBDT \citep{NIPS2017_6449f44a}. \textit{LightGBM} uses portion of the data with low memory cost applying two novel approaches for sampling: Gradient-based One-Side Sampling and Exclusive Feature Bundling. In contrast to \textit{XGBoost} where the trees are growth level-wise (horizontal), in \textit{LightGBM}, the decision trees are grown vertically, which can reduce more loss and provide a more accurate result.
 The goal of both algorithms, \textit{XGBoost} and \textit{LightGBM}, is the non-linear mapping from a set of input parameters to an outcome, namely, a prediction whose possible numerical values are spanned by the set of leaves. As an output, it provides a probability whether the source belongs to BL Lac or FSRQ sub-classes. 
\end{itemize}
\indent Both methods apply conceptually different approaches in transferring the input data into output models. The complex methods based on neural networks (like ANN) pose many challenges when applied to tabular data which contains sparsity (missing values, e.g., yearly fluxes in our case). The simple filling in of missing values with 0 or other constant might result in finding biased patterns in the data training or might significantly affect the obtained results. Instead, generally, the methods based on GBDT (like \textit{XGBoost} and \textit{LightGBM}) dominate when used on tabular data, showing superior performance. These algorithms can be trained on the data with missing values without doing imputation first and the tree branch directions for missing values are learned during training, each time deciding the best way to handle them. 
In addition, the algorithms based on decision trees are more applicable on the comparably small data sets (as in our case), as the complex methods based on neural networks tend to overfit the models. Thus, state-of-the-art algorithms, \textit{XGBoost} and \textit{LightGBM}, are more powerful and preferable tools for the classification of blazars. However, for a comparison we also performed classification using ANNs.
 
\subsection{Training and testing}
The data presented in previous section are used to train models and predict BCUs. Among the considered sources we dropped those which have three or more energy intervals with upper limits; in total 104 sources were dropped (73 BCUs and 31 FSRQs and BL Lacs). The entire dataset consists of 2219 rows from which 80\% was selected as training set while 20\% was the test set. We have used a 15-fold cross-validation procedure to evaluate the performance of the algorithms. In this procedure, the dataset is divided into 15 non-overlapping folds and the fitting is performed using 14 folds. Then, the model is validated using the remaining 15th fold. This procedure is repeated until every 15 folds serve as a validation and the average is taken as performance of the network. In this way, each of the 15 folds is given an opportunity to be used as a held-back validation set.

In our dataset each blazar (whether FSRQ or BL Lac) is characterized by 18 parameters: 12 yearly sorted \gray\ fluxes and 6 fluxes in each band. The BL Lacs and FSRQs with evident differences (see Section \ref{sources} and Fig. \ref{blazdistri}) occupy a different region in this parameter space and the goal is to quantify and determine the differences. In the ANN, the input neurons are 18, equal to the number of the input parameters, and we used three hidden layers with 64, 128 and 64 neurons. To prevent overfitting, we added two dropout layers, between hidden layers, to randomly set units to $0$ with a frequency of $0.4$ and $0.5$. The number of neurons in hidden layers was selected by reducing the number of neurons but keeping the performance accuracy. The \textit{XGBoost} and \textit{LightGBM} classifiers contain a set of important hyperparameters, among which the most important are the number of leaf nodes, the learning rate, and the number of iterations. Different values of these hyperparameters may increase the model performance, so their best values were found by hyperparameter tuning, i.e., the best version of the models are found by running many jobs that test a range of hyperparameters on the training and validation datasets. We used HyperOpt package \footnote{https://github.com/hyperopt/hyperopt} which uses a form of Bayesian optimization for parameter tuning. We found the following optimal parameters for \textit{LightGBM} model: the number of leaf nodes-31, learning rate-0.3, and number of iterations - 400, etc. However, in order to compare the resultant models, initially the model fitting was performed with the default parameters (\textit{XGBoost\_def} and \textit{LightGBM\_def}) and then with the optimized parameters (\textit{XGBoost\_opt} and \textit{LightGBM\_opt}). In all the trained models, the output was set up to have two possibilities, i.e., it returns the likelihood of a source to belong to either FSRQs ($L_{\rm FSRQ}$) or BL Lacs ($L_{\rm BL\:Lac}$). The likelihood is assigned such that $L_{\rm BL\:Lac}=1-L_{\rm FSRQ}$; the larger $L_{\rm BL\:Lac}$ (closer to 1) the higher the likelihood that the source is a BL Lac and vice-versa.
\begin{figure*}
    \centering
    \includegraphics[width=0.98\textwidth]{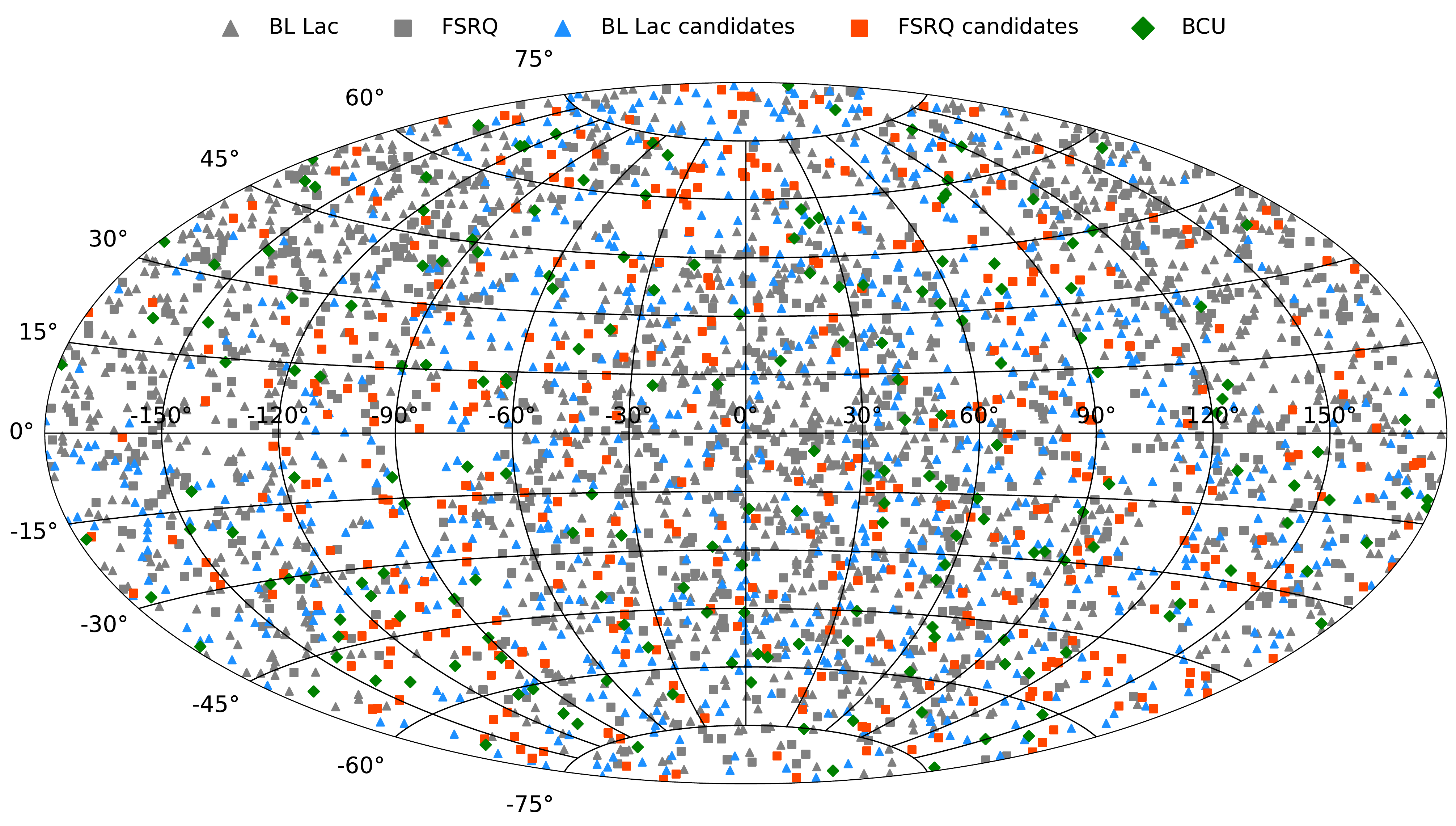}
    \caption{Hammer-Aitoff projection of FSRQs, BL Lacs and BCUs. The location of BL Lacs and FSRQs from 4FGL is shown with gray color while the new BL Lac and FSRQ candidates are in blue and orange, respectively.}
    \label{new_blaz}
\end{figure*}
\section{Results and Discussions}\label{res}
In this section, we discuss the results of classification of BCUs from 4FGL. Table \ref{aaa} provides summary results of applied models, showing their performance. To fully evaluate the effectiveness of the models, we compare the precision and recall. The precision measures the model's accuracy in classifying a sample as positive, i.e., the ability not to misclassify a BL Lac (FSRQ) as a FSRQ (BL Lac). Meanwhile, recall measures the model's ability to detect positive samples, i.e., the ability to identify all BL Lac (FSRQ) samples. Considering our dataset is slightly imbalanced, i.e., there are a disproportionate ratio of BL Lac and FSRQ classes (65:35), minority class (FSRQs in our case) has the highest interest from a learning point of view, as it can be under-classified. Therefore, we compare the models performance considering the recall of minority class (first column in Table \ref{aaa}) which shows that the \textit{LightGBM\_opt} model has the highest value for minority recall, $0.82$, and a satisfactory precision, $0.88$. Therefore, it provides the best BCU classification performance, so we report only the results obtained by this model.

In order to demonstrate the ability of \textit{LightGBM\_opt} to distinguish BL Lacs and FSRQs, the likelihood distribution of test sample is shown in Fig. \ref{likelihood}, left panel. The test sample contains 287 BL Lacs and 157 FSRQs, the ratio of which 287/157$\simeq$1.83 is the same as in the total sample (1436/783$\simeq$1.83). In the distribution, there are two evident and opposite peaks, BL Lacs (blue) centralized towards $L_{\rm BL\:Lac}=1$ while opposite for FSRQs $L_{\rm BL\:Lac}=0$, which clearly shows the ability of our model to separate BL Lacs and FSRQs from  the test sample (not used during the training). 

The performance of our applied model can be further examined from the confusion matrix shown in Fig. \ref{conf_matrix} which summarizes the number of true and predicted classes. The comparison of the number of correct (129-FSRQs and 261-BL Lacs) and incorrect (28-FSRQs and 26-BL Lacs) predictions shows that the model classifies BL Lacs and FSRQs perfectly. Other ways to analyze the effectiveness of the model are Receiver Operating Characteristic (ROC) and Precision-Recall curves shown in Fig. \ref{auc}. The ROC curve (left panel in Fig. \ref{auc}) represents the graph of the true positive rate versus the false positive ones and shows the performance of the model at all classification thresholds. The Area under the ROC Curve (AUC) is a measure of the model usefulness- the higher the AUC the better the model can distinguish between BL Lacs and FSRQs; in this case AUC=0.93. The Precision-Recall curve in Fig. \ref{auc} (right panel), the plot of the precision against the recall at a variety of thresholds, shows the trendoff between the precision and recall.

The accuracy of the network defined as positive association rate, i.e., how many BL Lacs (FSRQs) are correctly identified out of all BCUs, was optimized by selecting different classification thresholds for BL Lacs and FSRQs. The accuracy reaches 0.9 (90 \%) when the classification threshold of $L_{\rm BL\:Lac}>$ 0.44 identifies BL Lac candidates, while threshold $L_{\rm FSRQs}>$ 0.82 identifies FSRQ candidates.

We applied the best model to the entire 1420 BCU sample in 4FGL. It is found that 825 objects (58.1\%) have likelihood above the threshold of $L_{\rm BL\:Lac}=$ 0.44 and are classified as BL Lacs, 405 (28.5\%) are FSRQs having likelihood above $L_{\rm FSRQ}=$ 0.82, and 190 (13.3\%) remain unclassified.
The sky distribution of blazars locations in Galactic coordinates
and Hammer–Aitoff projection is shown in Fig. \ref{new_blaz}. The location of BL Lacs and FSRQs from 4FGL are shown with gray triangles and squares, respectively, most likely BL Lacs and FSRQs from BCUs are in blue triangles and orange squares, respectively, and the remaining unclassified BCUs are in green diamonds. 

\begin{table*}
\begin{center}
\caption{List of BCUs from 4FGL with the most significant association.}
\begin{tabular}{ccccccc} 
 \hline \hline
 Source Name & RA & Dec & Significance & Energy Flux & $L_{\rm BL\:Lac}$ & $L_{\rm FSRQ}$  \\
 \hline
4FGL J1925.0+2815 & 291.2683 &  28.2643 &         17.84 &  $(6.8\pm0.61)\times10^{-12}$ &    0.9989 &  0.0011 \\
4FGL J1958.1+2438 & 299.5284 &  24.6417 &         11.24 &  $(6.6\pm0.94)\times10^{-12}$ &    0.9986 &  0.0014 \\
4FGL J0121.7+5153 &  20.4389 &  51.8947 &          5.90 &  $(1.5\pm0.37)\times10^{-12}$ &    0.9983 &  0.0017 \\
4FGL J0538.6+0443 &  84.6630 &   4.7262 &          4.51 &  $(1.6\pm0.45)\times10^{-12}$ &    0.9982 &  0.0018 \\
4FGL J1401.1-3717 & 210.2971 & -37.2975 &         10.62 &  $(3.1\pm0.47)\times10^{-12}$ &    0.9980 &  0.0020 \\
4FGL J1846.7+7238 & 281.6849 &  72.6371 &          8.04 &  $(1.4\pm0.25)\times10^{-12}$ &    0.9978 &  0.0022 \\
4FGL J2142.1+4501 & 325.5291 &  45.0182 &          6.10 &  $(1.8\pm0.39)\times10^{-12}$ &    0.9976 &  0.0024 \\
4FGL J0215.3+7555 &  33.8285 &  75.9190 &          7.26 &  $(1.5\pm0.34)\times10^{-12}$ &    0.9975 &  0.0025 \\
4FGL J1535.3-3135 & 233.8391 & -31.5907 &          6.59 &  $(1.7\pm0.39)\times10^{-12}$ &    0.9973 &  0.0027 \\
4FGL J0606.5-4730 &  91.6416 & -47.5038 &         13.57 &  $(3.3\pm0.38)\times10^{-12}$ &    0.9972 &  0.0028 \\
4FGL J0507.4-3346 &  76.8591 & -33.7813 &         13.88 &  $(3.0\pm0.38)\times10^{-12}$ &    0.9971 &  0.0029 \\
4FGL J0954.2-2520 & 148.5681 & -25.3384 &          8.47 &  $(2.1\pm0.36)\times10^{-12}$ &    0.9971 &  0.0029 \\
4FGL J1943.6-0533 & 295.9249 &  -5.5665 &          4.86 &  $(2.2\pm0.53)\times10^{-12}$ &    0.9971 &  0.0029 \\
4FGL J1234.0-5735 & 188.5194 & -57.5961 &         29.72 & $(13.5\pm0.99)\times10^{-12}$ &    0.9970 &  0.0030 \\
4FGL J0213.8-6949 &  33.4704 & -69.8311 &          8.96 &  $(1.4\pm0.27)\times10^{-12}$ &    0.9970 &  0.0030 \\
4FGL J1412.0+3836 & 213.0130 &  38.6102 &          8.30 &  $(1.4\pm0.27)\times10^{-12}$ &    0.9970 &  0.0030 \\
4FGL J0830.1-0946 & 127.5427 &  -9.7728 &          7.53 &  $(1.8\pm0.36)\times10^{-12}$ &    0.9970 &  0.0030 \\
4FGL J1537.9-1344 & 234.4867 & -13.7335 &          5.76 &  $(1.6\pm0.40)\times10^{-12}$ &    0.9969 &  0.0031 \\
4FGL J2142.4+3659 & 325.6020 &  36.9856 &          9.83 &  $(2.8\pm0.43)\times10^{-12}$ &    0.9969 &  0.0031 \\
4FGL J1240.4-7148 & 190.1160 & -71.8156 &         21.15 &  $(7.7\pm0.61)\times10^{-12}$ &    0.9969 &  0.0031 \\
4FGL J0620.5-2512 &  95.1445 & -25.2129 &         17.24 &  $(7.8\pm0.70)\times10^{-12}$ &    0.0023 &  0.9977 \\
4FGL J0900.6-7408 & 135.1721 & -74.1440 &          8.25 &  $(3.6\pm0.54)\times10^{-12}$ &    0.0028 &  0.9972 \\
4FGL J2057.4-0723 & 314.3535 &  -7.3901 &          8.64 &  $(3.9\pm0.53)\times10^{-12}$ &    0.0033 &  0.9967 \\
4FGL J1830.2-4443 & 277.5504 & -44.7200 &         23.11 & $(10.0\pm0.72)\times10^{-12}$ &    0.0034 &  0.9966 \\
4FGL J0138.6+2923 &  24.6637 &  29.3855 &          7.20 &  $(2.5\pm0.46)\times10^{-12}$ &    0.0036 &  0.9964 \\
4FGL J0616.7-1049 &  94.1761 & -10.8230 &          6.02 &  $(5.1\pm1.11)\times10^{-12}$ &    0.0037 &  0.9963 \\
4FGL J0732.7-4638 & 113.1774 & -46.6488 &         11.17 &  $(4.9\pm0.97)\times10^{-12}$ &    0.0037 &  0.9963 \\
4FGL J0953.1-3005 & 148.2779 & -30.0979 &         10.08 &  $(3.5\pm0.51)\times10^{-12}$ &    0.0041 &  0.9959 \\
4FGL J0841.0-2744 & 130.2630 & -27.7468 &          6.09 &  $(2.6\pm0.54)\times10^{-12}$ &    0.0043 &  0.9957 \\
4FGL J0348.8+4610 &  57.2185 &  46.1695 &          6.51 &  $(3.6\pm0.94)\times10^{-12}$ &    0.0046 &  0.9954 \\
4FGL J2139.9+3910 & 324.9929 &  39.1711 &          3.72 &  $(2.6\pm0.67)\times10^{-12}$ &    0.0050 &  0.9950 \\
4FGL J0008.0-3937 &   2.0048 & -39.6320 &          5.44 &  $(2.1\pm0.41)\times10^{-12}$ &    0.0052 &  0.9948 \\
4FGL J1437.3-3239 & 219.3259 & -32.6569 &          5.17 &  $(1.7\pm0.53)\times10^{-12}$ &    0.0053 &  0.9947 \\
4FGL J0118.7-0848 &  19.6884 &  -8.8080 &          7.56 &  $(2.7\pm0.43)\times10^{-12}$ &    0.0053 &  0.9947 \\
4FGL J2141.7-6410 & 325.4305 & -64.1792 &         65.83 & $(25.0\pm0.78)\times10^{-12}$ &    0.0054 &  0.9946 \\
4FGL J1821.6+6819 & 275.4034 &  68.3242 &         34.47 & $(12.1\pm0.86)\times10^{-12}$ &    0.0055 &  0.9945 \\
4FGL J0429.0-0006 &  67.2549 &  -0.1006 &          3.39 &  $(2.3\pm0.63)\times10^{-12}$ &    0.0055 &  0.9945 \\
4FGL J0501.0-2423 &  75.2732 & -24.3935 &          5.11 &  $(3.5\pm0.72)\times10^{-12}$ &    0.0059 &  0.9941 \\
4FGL J2318.2+1915 & 349.5568 &  19.2560 &         13.17 &  $(6.2\pm0.56)\times10^{-12}$ &    0.0059 &  0.9941 \\
4FGL J1421.6-4819 & 215.4125 & -48.3317 &          6.61 &  $(3.8\pm0.85)\times10^{-12}$ &    0.0059 &  0.9941 \\
\end{tabular}
\label{tsources}
\end{center}

\end{table*}

The likelihood distribution of the model applied to BCU sample is shown on the right panel of Fig. \ref{likelihood}. As expected, it mimics the same trend as seen for BL Lacs and FSRQs in test sample (left panel of Fig. \ref{likelihood}). The ratio of BL Lacs to FSRQs identified from BCUs is 2.03, similar to the result obtained in \citet{2020MNRAS.493.1926K} which classified BCUs in Fermi-LAT 8-year source catalogue using ANNs. Table \ref{tsources} shows the portion of the sources having higher probability of being BL Lacs and FSRQs; for each source the name in the catalogue, RA and Dec, detection significance ($\sigma$), energy flux and its uncertainty, and the probabilities to be BL Lac or FSRQ are reported. As it can be seen from the table, the used algorithm can classify BCUs as BL Lacs or FSRQs with a high probability.
However, there are also sources with intermediate likelihoods falling in the region where the two sub-classes overlap, so they cannot be classified by the model. The full BCUs classification table that includes also the sources which have lower association likelihood is available in the online supplementary material and at the following github repository \url{https://github.com/mherkhachatryan/BCU-Classification.git}.
\begin{figure*}
    \centering
    \includegraphics[width=0.49\textwidth]{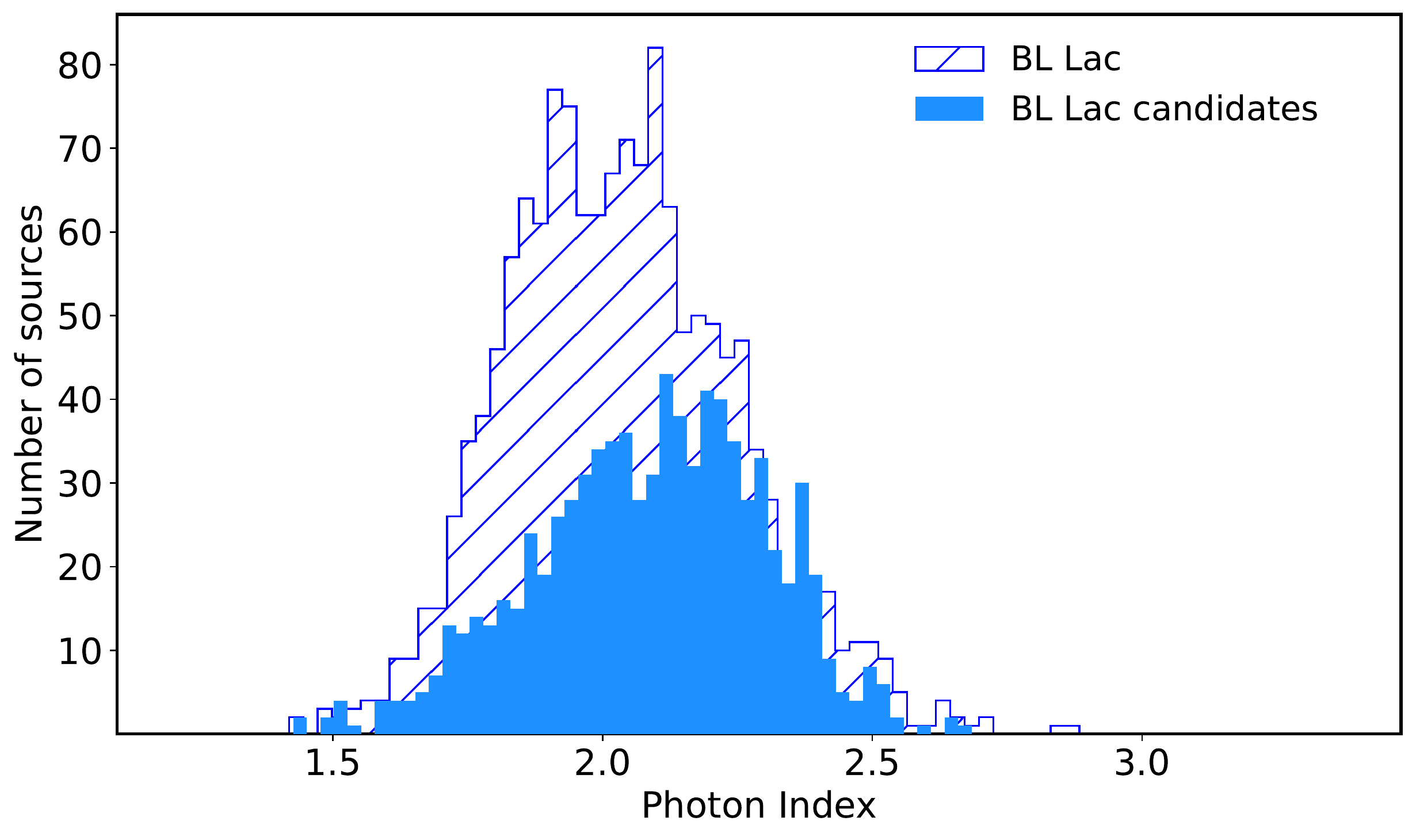}
    \includegraphics[width=0.49\textwidth]{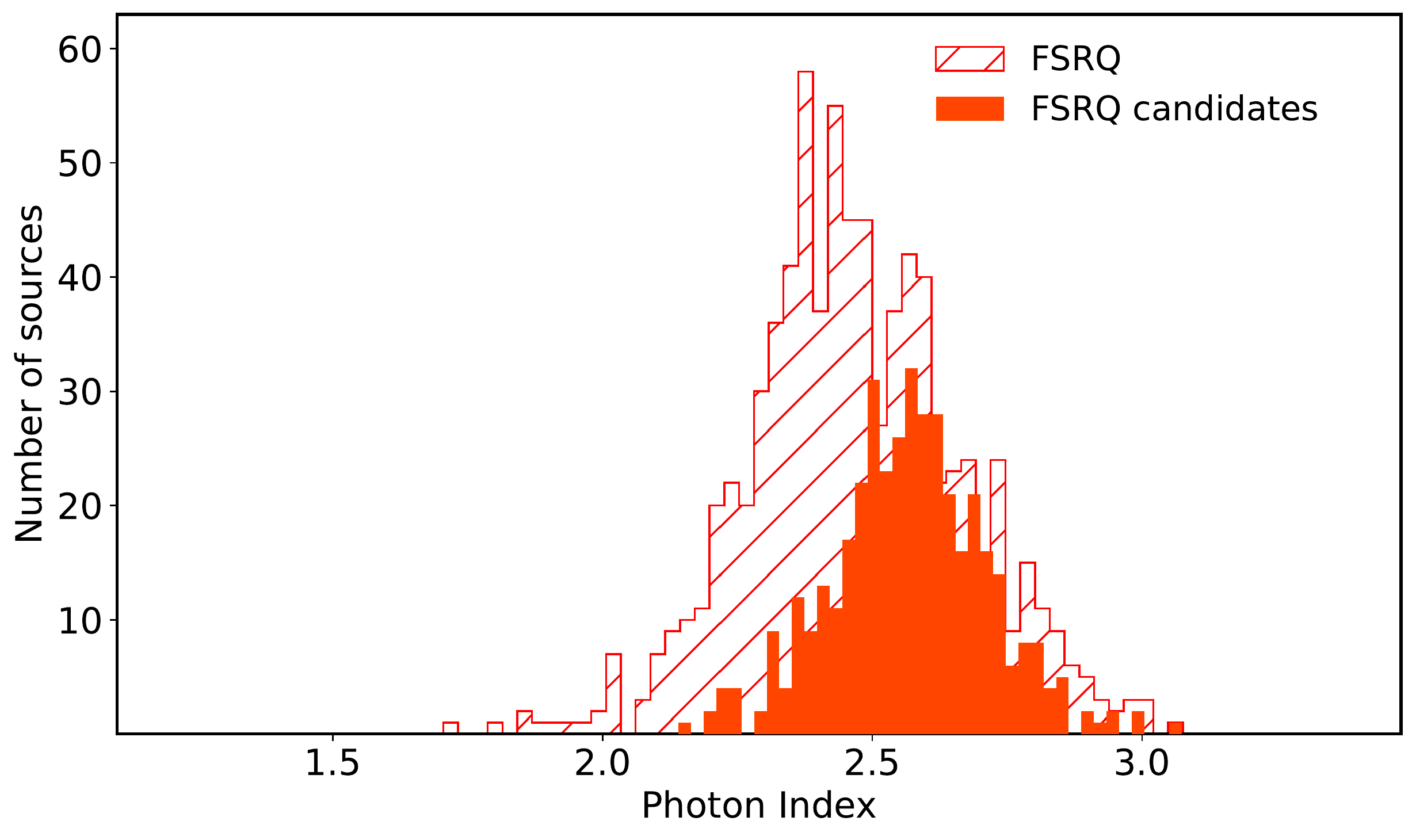}
    \caption{The power-law photon index distribution of BL Lacs (left) and FSRQs (right) in 4FGL. The distributions are compared with newly classified BL Lac and FSRQ candidates (filled histograms).}
    \label{PL_dist}
\end{figure*}

The comparison of BL Lac and FSRQ candidate list presented here with the previous studies is not straightforward and rather difficult as the number of blazars and their classifications are changing in different versions of the catalogues. However, to show the accuracy of our method, we present a general comparison of our results with those presented in \citet{2022JCAP...04..023B} where the BCUs from 4FGL DR2 (a catalogue version preceding 4FGL DR3) are classified using Bayesian neural networks. Applying tight selection criteria, their list contains  429 BL Lacs and 178 FSRQs. Among their BL Lac candidates 355 (82.7\%) objects are in agreement with our prediction and we found a difference in 74. Among those 74, in 4FGL DR3 53 objects have already been classified as BL Lacs, 2 as FSRQs and 1 as normal galaxy; 13 objects were dropped by us because of insufficient spectral data (see Section \ref{model}). So disagreement is found only with 5 objects: all remained unclassified. Similarly, out of 178 FSRQs in their list, 148 (83.1\%) match with our results; from the remaining 31 sources in 4FGL DR3 4 have already been classified as FSRQs  and 2 are missing, we excluded 6 objects, 13 objects remained unclassified with a probability between $L_{\rm BL\:Lac}=0.18-0.44$ and we found disagreement for 5 objects. Similar picture can be drawn when comparing with the BL Lac and FSRQ list by applying loose selection criteria. Thus, our results are in a good agreement with those presented in \citet{2022JCAP...04..023B} and obtained by a different method.
\subsection{BL Lac and FSRQ candidates versus BL Lacs and FSRQs}

The idea of the classification presented in the previous subsection is to identify new BL Lac and FSRQ candidates among the unclassified blazars. In this subsection, we compare and contrast the properties of newly identified and known sources.

The spectral difference between BL Lacs and FSRQs in 0.1-300 GeV band is well known. The distribution of power-law photon indexes of BL Lacs and FSRQs from 4FGL are shown in Fig. \ref{PL_dist} left and right panels, respectively (blue and orange shaded areas). The mean and standard deviation of these distributions is $2.03\pm0.21$ for BL Lacs and $2.47\pm0.20$ for FSRQs; on the average, BL Lacs spectra are harder than those of FSRQs. The distribution of newly classified BL Lac and FSRQ candidates is shown by filled blue and orange areas in the left and right panels of Fig. \ref{PL_dist}, respectively. The distribution of likely BL Lacs and FSRQs is centered on $2.09\pm0.21$ and $2.57\pm0.14$, respectively, in an excellent agreement with the distributions of known BL Lacs and FSRQs. Thus, BL Lacs and FSRQs classified from BCUs have similar spectral characteristics as compared with those derived for known sources. As the power-law index was not considered in the model training, this comparison is an effective way to illustrate the power of our method.
\begin{figure}
    \centering
    \includegraphics[width=0.45\textwidth]{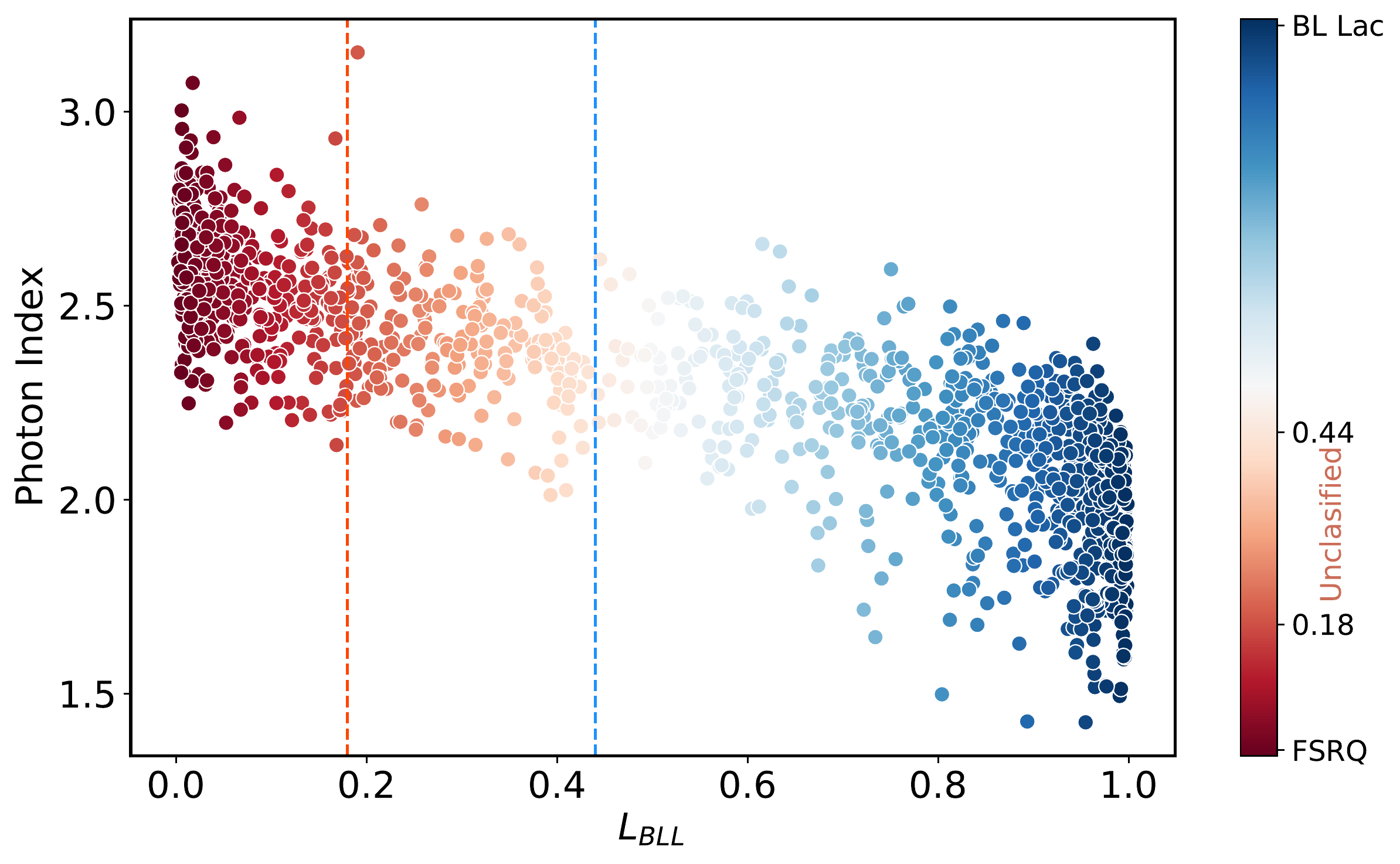}
    \caption{Photon index versus the probability. The color bar shows the probabilities of source associations: FSRQs are in dark red and BL Lacs in dark blue.}
    \label{pl_lik}
\end{figure}

Another way to visualize the power-law photon index difference between BL Lacs and FSRQs is to plot the photon index versus $L_{\rm BL\:Lac}$ (Fig. \ref{pl_lik}). There is a clear correlation between the photon index and probability; higher $L_{\rm BL\:Lac}$ corresponds to lower index and vice-versa. This very well follows the spectral trend observed for BL Lacs and FSRQs, namely higher $L_{\rm BL\:Lac}$ (more likely BL Lacs) corresponds to harder spectra while FSRQs (lower $L_{\rm BL\:Lac}$) appear with softer spectra. 
\begin{figure}
    \centering
    \includegraphics[width=0.45\textwidth]{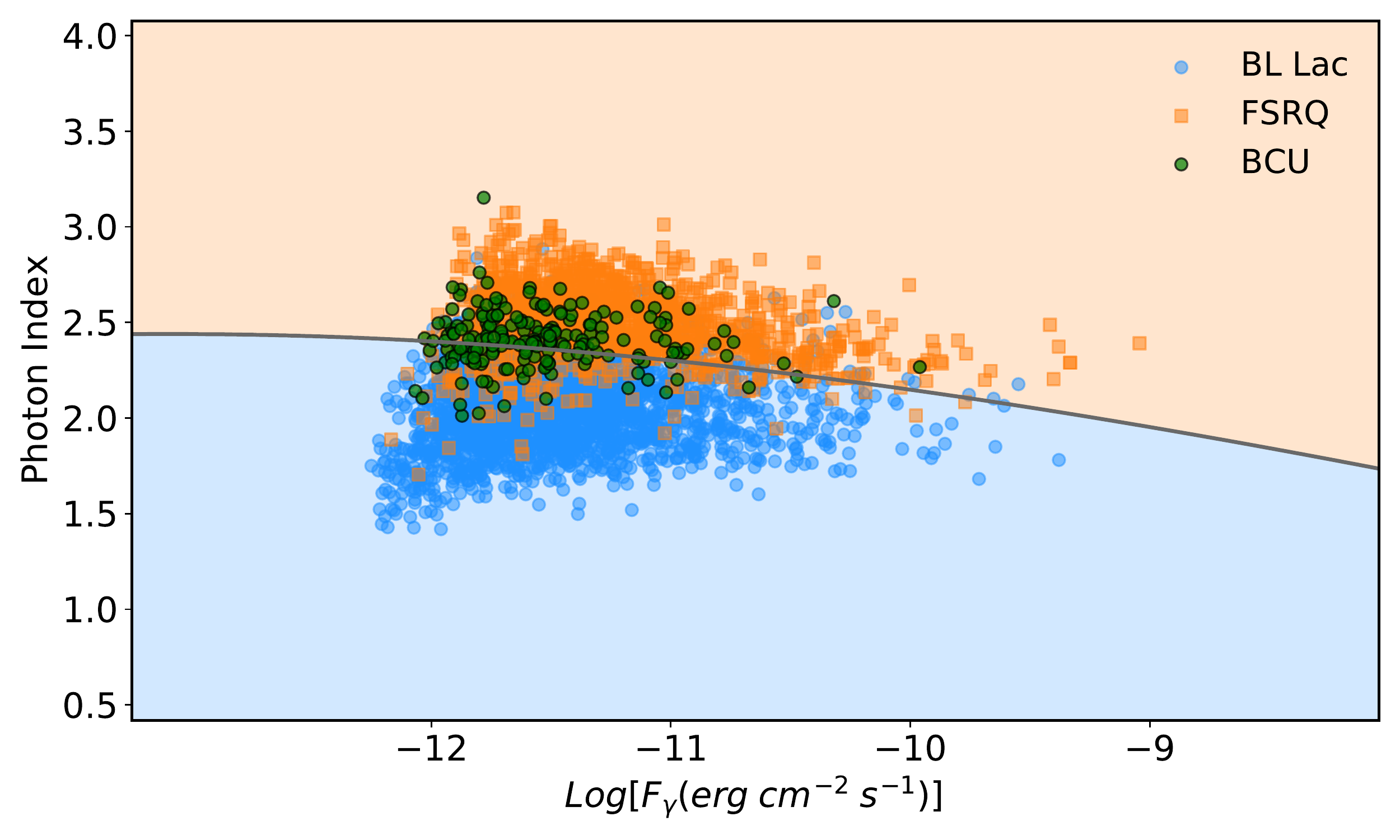}
    \caption{Photon index versus energy flux above 100 MeV. The curve represents the approximate boundary of two sub-classes separation.}
    \label{fl_pl}
\end{figure}

Vertical blue and orange dashed lines show the classification threshold defined for BL Lacs and FSRQs. 57.2 \% of total BCUs are classified as BL Lacs (right region from the blue dashed line in Fig. \ref{pl_lik}) and 28.9 \% are FSRQs (left region from the orange dashed line in Fig. \ref{pl_lik}). 
The sources falling between the orange and blue dashed lines (13.9 \%) remain unclassified, their power-law index is in the range defined for BL Lacs and FSRQs.

The increased number of BL Lacs and FSRQs ($2261$ and $1188$, respectively) allows to compare their properties with improved statistics. A convenient way to compare the properties of different blazar classes is through plotting the power-law photon index versus the flux (Fig. \ref{fl_pl}). Since the spectra of some sources deviate from the simple power-law model, we computed the energy flux between 100 MeV and 100 GeV using the power-law model parameters given in 4FGL, namely, reported flux density, pivot energy and power-law photon index. The BL Lacs and FSRQs both from 4FGL and classified from BCUs are shown in blue and orange, respectively and the remaining unclassified BCUs are in green. The reported energy fluxes vary from $5.62\times10^{-13}\:{\rm erg\:cm^{-2}\:s^{-1}}$ to $9.06\times10^{-10}\:{\rm erg\:cm^{-2}\:s^{-1}}$ while the photon index is in the range from $1.42$ to $3.08$. 
A Kolmogorov–Smirnov test gives a probability of 0.31 that FSRQs and
FSRQ candidates come from the same parent population and the probability is 0.15 for the BL Lacs and BL Lac candidates.
From the distribution it is possible to quantify the space occupied by BL Lacs and FSRQs in the photon index versus energy flux plane. The boundary between these two classes (decision boundary) was found by using Gaussian naive Bayes classification. The black line in Fig. \ref{fl_pl} is the decision boundary, which corresponds to the curve that optimally separates the two classes. In other words, by computing the probabilities the algorithm optimally divides the plane in a such way as to have the highest number of BL Lacs and FSRQs below and above the line, respectively. The boundary line defines $\simeq2.41$, $2.31$ and $2.15$ indices at $10^{-12}\:{\rm erg\:cm^{-2}\:s^{-1}}$, $10^{-11}\:{\rm erg\:cm^{-2}\:s^{-1}}$ and $10^{-10}\:{\rm erg\:cm^{-2}\:s^{-1}}$ fluxes, respectively, well separating the two-classes: 91.2 \% of BL Lacs occupy the region below the line while 86.1 \% of FSRQs are above. The remaining BCUs (green circles in Fig. \ref{fl_pl}) are distributed above and below the limit; 65.1\% of BCUs occupy the space more characteristic for FSRQs, while 34.9 \% show properties more similar to BL Lacs.
The limit presented in Fig. \ref{fl_pl}, which is a physical distinction between the two classes of blazars, BL Lacs and FSRQs, based on photon index and flux, was obtained using a large number of BL Lacs and FSRQs, 3449 in total. The accuracy of Gaussian naive Bayes classification is 90\% implying the line satisfactorily well separates the two classes of blazars. Even if this is not a strict limit, it still can be used as a reference limit for BL Lac and FSRQ division.
\begin{figure*}
    \centering
    \includegraphics[width=0.45\textwidth]{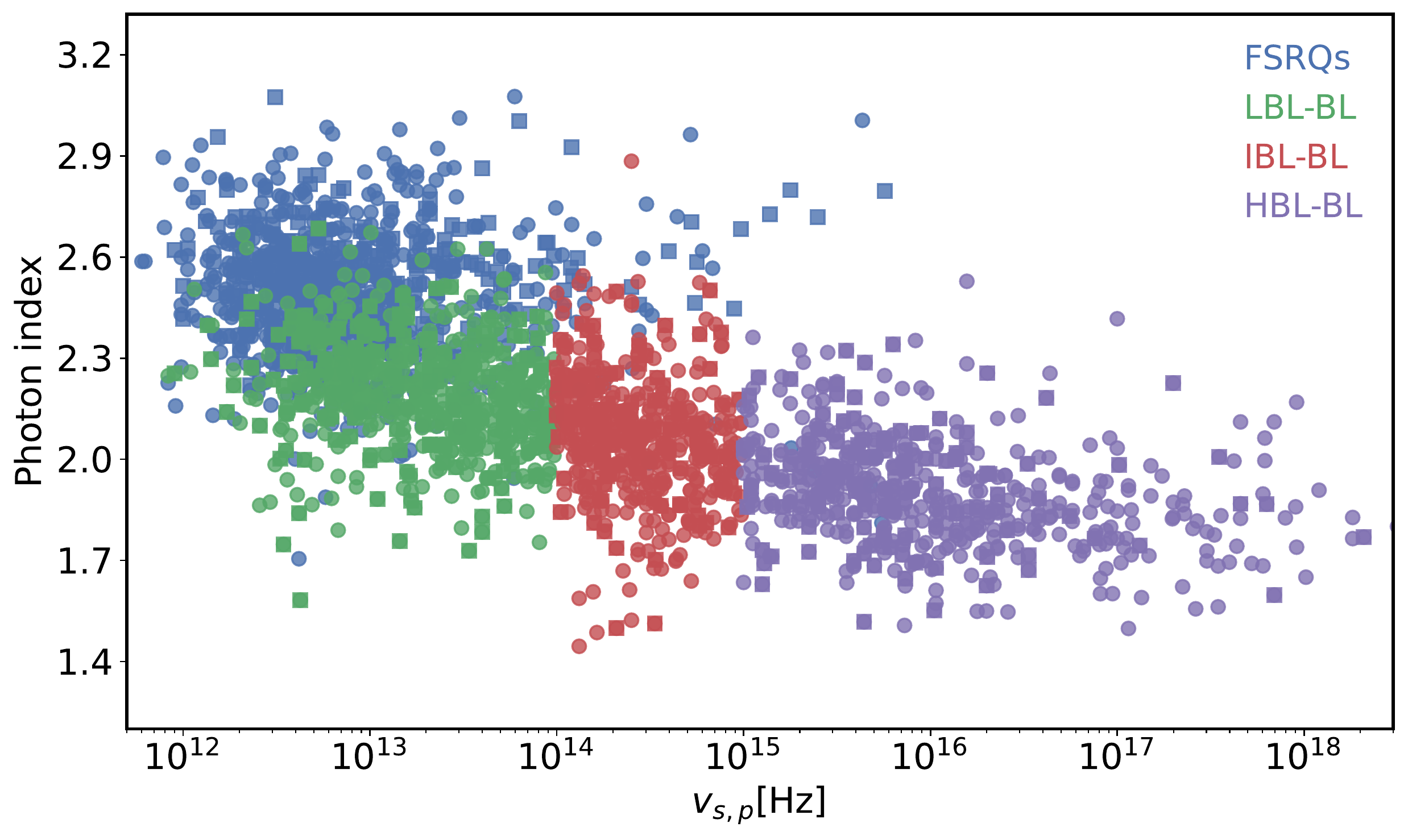}
    \includegraphics[width=0.45\textwidth]{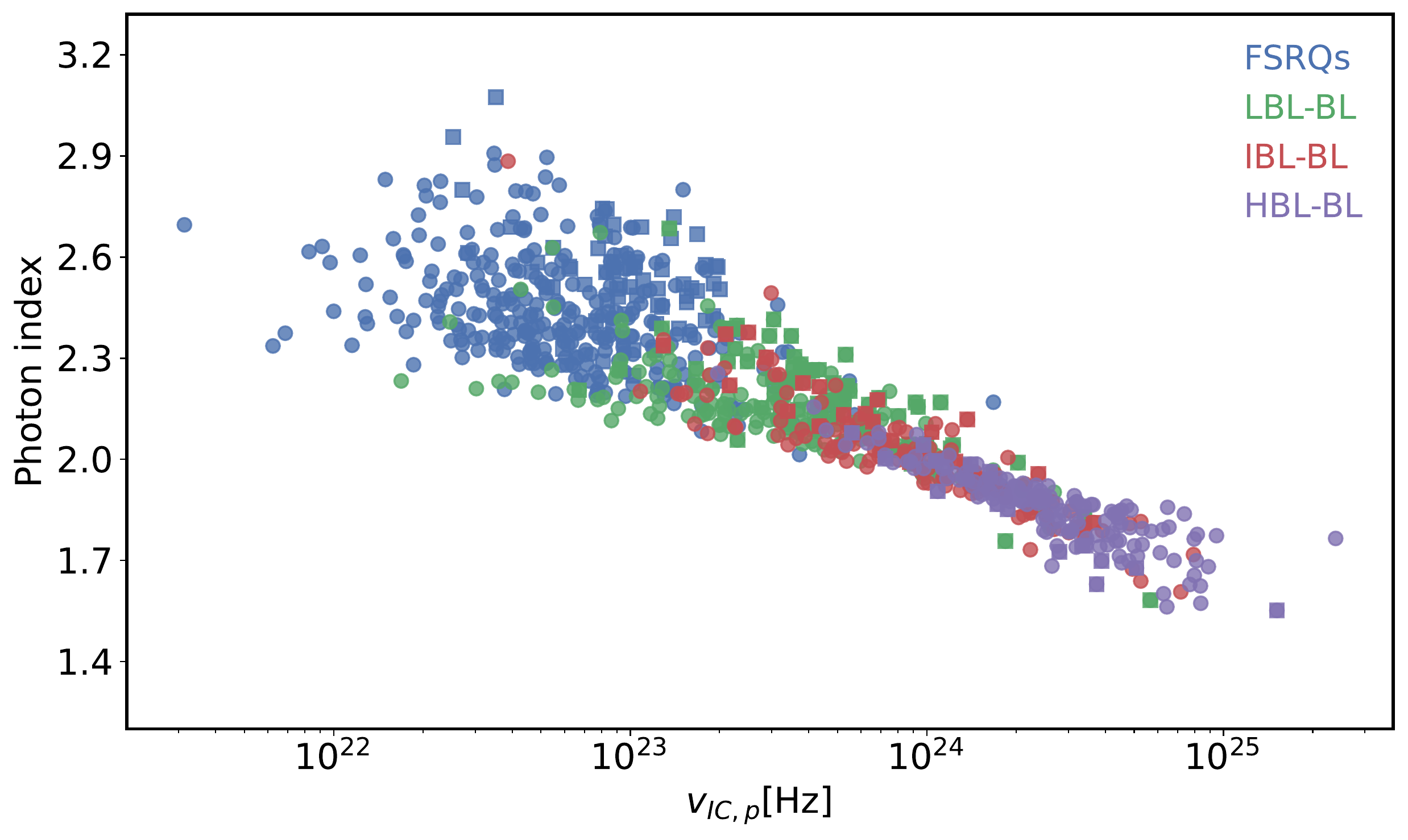}
    \caption{ {\it Left panel:} \gray\ photon index versus the synchrotron peak frequency. {\it Right panel:}  \gray\ photon index versus the inverse Compton peak frequency. The FSRQs, LBLs, IBLs and HBLs are in blue, green, red and purple, respectively. BL Lacs and FSRQs from 4FGL are shown with circles, while BL Lac and FSRQ candidates are with squares.}
    \label{fl_peaks}
\end{figure*}

It is also interesting to compare the photon index with the frequency of the synchrotron peak ($\nu_{\rm s, p}$). In the previous studies, a strong anticorrelation between these two parameters was already reported \citep[e.g.,][]{2015ApJ...810...14A, 2020ApJ...892..105A}. Now, with updated number of BL Lacs and FSRQs tighter constraints on the photon index distribution on different blazar types can be obtained \footnote{We note that for a more meaningful comparison, $\nu_{\rm s, p}$ should be corrected by a factor of $1+z$ but this will reduce the source sample size, as $z$ is not measured for many sources.}. Fig. \ref{fl_peaks} (left panel) shows the photon index versus $\nu_{\rm s, p}$ where BL Lacs are separated into LBL, IBL and HBL classes. We note that also FSRQs are mostly LBLs, but we show and discuss their properties by separating them from BL Lacs that are classified as LBLs, IBLs and HBLs. For the FSRQs (blue), LBLs (green), IBLs (red) and HBLs (purple) the mean and rms of the photon index are $2.49\pm0.18$, $2.20\pm0.16$, $2.06\pm0.18$ and $1.90\pm0.16$, respectively. As expected, the distributions of FSRQs and LBLs are similar but LBLs have a slightly harder \gray\ photon index,  but, clearly, as compared with them  IBLs and HBLs occupy a different region in the \gray\ photon index $\nu_{\rm s, p}$ plane; the \gray\ photon index becomes smaller (harder) from FSRQs to HBLs, $\Gamma_{\rm \gamma, FSRQs}>\Gamma_{\rm \gamma, LBLs}> \Gamma_{\rm \gamma, IBLs} > \Gamma_{\rm \gamma, HBLs}$.

The distribution of the high energy peak frequency (referred as inverse-Compton peak $\nu_{\rm IC, p}$) for the considered sources is shown in Fig. \ref{fl_peaks} (right panel). This is a new parameter available in the fourth catalog of AGNs detected by Fermi LAT - Data Release 3 \citep[][]{2022arXiv220912070T} which has been estimated by fitting the significantly curved spectrum with a log-parabolic model. For the current study, we excluded all the sources for which the uncertainty on the high energy peak estimation is large (requiring the value to be larger than $1.5\times$ error) which resulted in 893 blazars with measured $\nu_{\rm IC, p}$, among which 369 are FSRQs, 217 are LBLs, 143 are IBLs and 164 are HBLs. The distributions of LBLs, IBLs and HBLs overlap and are slightly separated from FSRQs toward higher frequencies. The fit of the linear function $\Gamma_{\gamma}=\Gamma_0+\alpha\times \nu_{\rm IC, p}$ yields $\alpha=-0.21\pm0.02$ and $\alpha=-0.31\pm0.01$ for FSRQs and BL Lacs (considering all LBLs, IBLs and HBLs), respectively. This shows that $\Gamma_{\gamma}$ of BL Lacs becomes steeper with increasing $\nu_{\rm IC, p}$.

\begin{figure}
    \centering
    \includegraphics[width=0.45\textwidth]{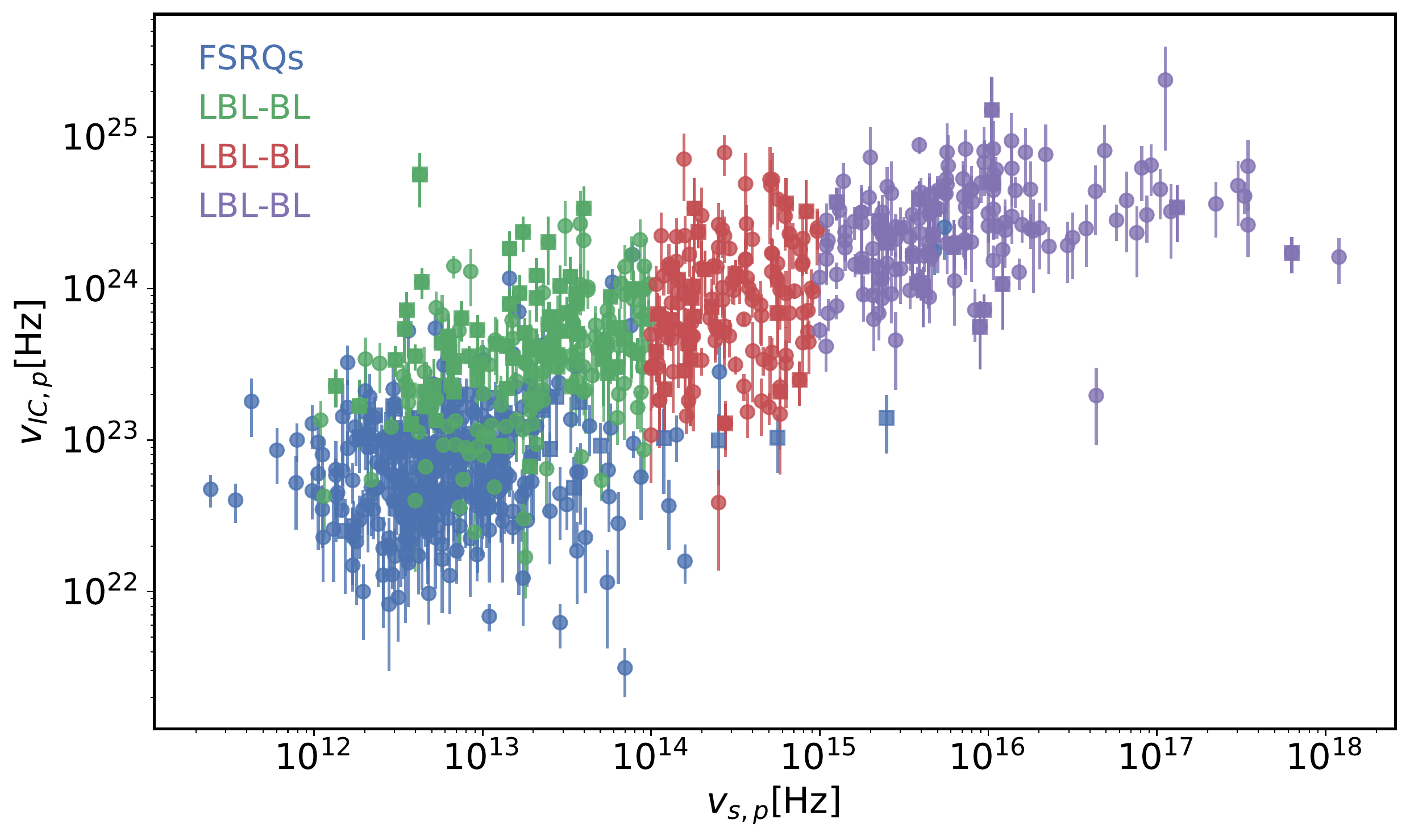}
    \caption{The synchrotron peak frequency ($\nu_{\rm s, p}$) as a function of the high energy peak frequency ($\nu_{\rm IC, p}$). The same color code and plot markers as in Fig. \ref{fl_peaks}.}
    \label{fl_he_peak}
\end{figure}

The comparison of synchrotron ($\nu_{\rm s, p}$) and high energy ($\nu_{\rm IC, p}$) peak frequencies estimated for different blazar sub-classes is shown in Fig. \ref{fl_he_peak}. In general, the $\nu_{\rm IC, p}$ of FSRQs is at lower frequencies than those of BL Lacs which in their turn show different tendencies for LBLs, IBLs and HBLs. $\nu_{\rm IC, p}$ increases along with the increase of $\nu_{\rm s, p}$; the linear fit ($\nu_{\rm IC, p}=\nu_{0}+\kappa\times \nu_{\rm s, p}$) for BL Lacs shows a slope of $\kappa=0.33\pm0.01$. Only $10.1\%$ of LBLs have $\nu_{\rm IC, p}$ above $10^{24}$ Hz, while the percentage is $41.2\%$ for IBLs and $89.0\%$ for HBLs. The difference between the FSRQs and BL Lacs as well as between LBLs, IBLs and HBLs is expected from simple theoretical considerations. Within a simple one-zone synchrotron/synchrotron-self Compton (SSC) model, which is successfully applied to model the SEDs of BL Lacs \citep[e.g., see][]{1985A&A...146..204G}, assuming Thomson regime for the inverse Compton scattering, $\nu_{\rm s, p}$ and $\nu_{\rm IC, p}$ are linked by $\nu_{\rm IC, p}/\nu_{\rm s, p}=4/3 (\gamma_{\rm p}^{\rm SSC})^2$. So, it is natural that for the sources with higher synchrotron peak (HBLs) also $\nu_{\rm IC, p}$ is at higher frequencies. However, the $\nu_{\rm s, p}$ and $\nu_{\rm IC, p}$ relation is not valid for the FSRQs, where the high energy emission is most likely due to inverse Compton scattering of external photons (UV and IR external radiation fields which usually dominate over the jet synchrotron emission), so the $\nu_{\rm IC, p}$ scales with the average energy of up-scattered photons and bulk Lorentz factor of the emitting region.

\section{Conclusions}\label{conc}
In this paper, we performed machine learning classification of blazar candidates of uncertain type. By training and constructing predictive models which forecast the likelihood of a source to belong to a particular class of blazars, the classification of BCUs based on their direct observable spectral and temporal properties in the \gray\ band is conducted.

The models were trained on the set of \gray\ parameters (spectra and light curves) of BL Lacs and FSRQs from the latest and most detailed \gray\ catalog (4FGL DR3) which is based on accumulation of data in twelve years (2008-2022).
Different machine learning algorithms were applied to classify blazars by dividing the entire data set into 80\% and 20\% train and test subsets and performing 15 fold cross validation. The algorithms based on gradient-boosted decision trees are preferred and outperform the other models because the available data set is comparatively small and contains missing data points. As a result, \textit{LightGBM}- a state-of-the-art classification model based on gradient boosted trees shows the highest performance with a weighted recall of 0.88 and precision of 0.88.

The best model was applied to 1420 BCUs included in 4FGL to obtain the probability of their association to one of the blazar sub-classes and address the question of their nature. As a result, among the BCUs 825 (58.1\%) are BL Lac candidates, 405 (28.5\%) are FSRQ candidates and only 190 (13.3\%) cannot be classified by our model. The \gray\ spectral properties (e.g., power-law photon index) of already classified and BL Lac and FSRQ candidates are in an excellent agreement, showing the validity of our model.

The results of BCU classification reported here although cannot conclusively give the type of a BCU but they can be useful for statistical population studies or for planing optical monitoring of blazars. For example, the distributions of BL Lacs and FSRQs from a more complete list show clustering in the photon index and energy flux plane, clearly separating those two sub-classes. According to this criteria, the majority of BCUs (65.1\%) that remained unclassified in our model show properties more similar to FSRQs. For a larger blazar sample, the distribution of the synchrotron peak frequency ($\nu_{\rm s, p}$) versus the \gray\ photon index confirms the strong difference between FSRQs and BL Lacs as well as between LBLs, IBLs and HBLs. In the distribution of the high energy peak frequency versus the \gray\ photon index most of the FSRQs occupy the region of $\Gamma_{\rm \gamma}>2.3$ and $\nu_{\rm IC, p}<10^{23}$ Hz whereas $\nu_{\rm IC, p}$ of BL Lacs can reach higher frequencies with harder \gray\ photon index, however no distinction between LBLs, IBLs and HBLs sublasses is possible. Instead, the comparison of $\nu_{\rm s, p}$ and $\nu_{\rm IC, p}$ shows a remarkable difference between FSRQs and BL Lacs as well as between LBLs, IBLs and HBLs. The BL Lacs have much larger $\nu_{\rm IC, p}$ with a mean of $1.6\times10^{24}$ Hz as compared with that of FSRQS with $1.1\times10^{23}$; $43.3\%$ of BL Lacs have a peak above $10^{24}$ Hz. Among BL Lacs, HBLs have higher $\nu_{\rm IC, p}$ (for $89.0\%$ of HBLs $\nu_{\rm IC, p}\geq10^{24}$ Hz) whereas it is lower for LBLs; the linear fit shows a slope of $0.33\pm0.01$, so $\nu_{\rm IC, p}$ increases with $\nu_{\rm s, p}$. Such a correlation is in agreement with expectations from one-zone synchrotron-self-Compton scenarios.
\section*{Acknowledgements}
We thank the anonymous referee for constructive comments.

This work was supported by the Science Committee of the Republic of Armenia, in the frames of the research project No 20TTCG-1C015.
\section*{Data availability}
The data underlying this article will be shared on reasonable request to the corresponding author. The table of the complete list of BCU classifications is available at MNRAS online as supplementary material.



\bibliographystyle{mnras}
\bibliography{biblio} 

\begin{thebibliography}{}
\makeatletter
\relax
\def\mn@urlcharsother{\let\do\@makeother \do\$\do\&\do\#\do\^\do\_\do\%\do\~}
\def\mn@doi{\begingroup\mn@urlcharsother \@ifnextchar [ {\mn@doi@}
  {\mn@doi@[]}}
\def\mn@doi@[#1]#2{\def\@tempa{#1}\ifx\@tempa\@empty \href
  {http://dx.doi.org/#2} {doi:#2}\else \href {http://dx.doi.org/#2} {#1}\fi
  \endgroup}
\def\mn@eprint#1#2{\mn@eprint@#1:#2::\@nil}
\def\mn@eprint@arXiv#1{\href {http://arxiv.org/abs/#1} {{\tt arXiv:#1}}}
\def\mn@eprint@dblp#1{\href {http://dblp.uni-trier.de/rec/bibtex/#1.xml}
  {dblp:#1}}
\def\mn@eprint@#1:#2:#3:#4\@nil{\def\@tempa {#1}\def\@tempb {#2}\def\@tempc
  {#3}\ifx \@tempc \@empty \let \@tempc \@tempb \let \@tempb \@tempa \fi \ifx
  \@tempb \@empty \def\@tempb {arXiv}\fi \@ifundefined
  {mn@eprint@\@tempb}{\@tempb:\@tempc}{\expandafter \expandafter \csname
  mn@eprint@\@tempb\endcsname \expandafter{\@tempc}}}

\bibitem[\protect\citeauthoryear{{Abdollahi} et~al.,}{{Abdollahi}
  et~al.}{2020}]{2020ApJS..247...33A}
{Abdollahi} S.,  et~al., 2020, \mn@doi [\apjs] {10.3847/1538-4365/ab6bcb},
  \href {https://ui.adsabs.harvard.edu/abs/2020ApJS..247...33A} {247, 33}

\bibitem[\protect\citeauthoryear{{Abdollahi} et~al.,}{{Abdollahi}
  et~al.}{2022}]{2022ApJS..260...53A}
{Abdollahi} S.,  et~al., 2022, \mn@doi [\apjs] {10.3847/1538-4365/ac6751},
  \href {https://ui.adsabs.harvard.edu/abs/2022ApJS..260...53A} {260, 53}

\bibitem[\protect\citeauthoryear{{Ackermann} et~al.,}{{Ackermann}
  et~al.}{2012}]{2012ApJ...753...83A}
{Ackermann} M.,  et~al., 2012, \mn@doi [\apj] {10.1088/0004-637X/753/1/83},
  \href {https://ui.adsabs.harvard.edu/abs/2012ApJ...753...83A} {753, 83}

\bibitem[\protect\citeauthoryear{{Ackermann} et~al.,}{{Ackermann}
  et~al.}{2015}]{2015ApJ...810...14A}
{Ackermann} M.,  et~al., 2015, \mn@doi [\apj] {10.1088/0004-637X/810/1/14},
  \href {https://ui.adsabs.harvard.edu/abs/2015ApJ...810...14A} {810, 14}

\bibitem[\protect\citeauthoryear{{Ajello} et~al.,}{{Ajello}
  et~al.}{2020}]{2020ApJ...892..105A}
{Ajello} M.,  et~al., 2020, \mn@doi [\apj] {10.3847/1538-4357/ab791e}, \href
  {https://ui.adsabs.harvard.edu/abs/2020ApJ...892..105A} {892, 105}

\bibitem[\protect\citeauthoryear{{Ajello} et~al.,}{{Ajello}
  et~al.}{2022}]{2022arXiv220912070T}
{Ajello} M.,  et~al., 2022, \mn@doi [\apjs] {10.3847/1538-4365/ac9523}, \href
  {https://ui.adsabs.harvard.edu/abs/2022ApJS..263...24A} {263, 24}

\bibitem[\protect\citeauthoryear{{Arsioli} \& {Dedin}}{{Arsioli} \&
  {Dedin}}{2020}]{2020MNRAS.498.1750A}
{Arsioli} B.,  {Dedin} P.,  2020, \mn@doi [\mnras] {10.1093/mnras/staa2449},
  \href {https://ui.adsabs.harvard.edu/abs/2020MNRAS.498.1750A} {498, 1750}

\bibitem[\protect\citeauthoryear{{Atwood} et~al.,}{{Atwood}
  et~al.}{2009}]{2009ApJ...697.1071A}
{Atwood} W.~B.,  et~al., 2009, \mn@doi [\apj] {10.1088/0004-637X/697/2/1071},
  \href {https://ui.adsabs.harvard.edu/abs/2009ApJ...697.1071A} {697, 1071}

\bibitem[\protect\citeauthoryear{Auld, Bridges, Hobson  \& Gull}{Auld
  et~al.}{2007}]{10.1111/j.1745-3933.2006.00276.x}
Auld T.,  Bridges M.,  Hobson M.~P.,   Gull S.~F.,  2007, \mn@doi [Monthly
  Notices of the Royal Astronomical Society: Letters]
  {10.1111/j.1745-3933.2006.00276.x}, 376, L11

\bibitem[\protect\citeauthoryear{{Bhat} \& {Malyshev}}{{Bhat} \&
  {Malyshev}}{2022}]{2022A&A...660A..87B}
{Bhat} A.,  {Malyshev} D.,  2022, \mn@doi [\aap] {10.1051/0004-6361/202140766},
  \href {https://ui.adsabs.harvard.edu/abs/2022A&A...660A..87B} {660, A87}

\bibitem[\protect\citeauthoryear{Bishop}{Bishop}{1995}]{bishop1995neural}
Bishop C.,  1995, {Neural networks for pattern recognition}.
Oxford University Press, USA

\bibitem[\protect\citeauthoryear{{Butter}, {Finke}, {Keil}, {Kr{\"a}mer}  \&
  {Manconi}}{{Butter} et~al.}{2022}]{2022JCAP...04..023B}
{Butter} A.,  {Finke} T.,  {Keil} F.,  {Kr{\"a}mer} M.,   {Manconi} S.,  2022,
  \mn@doi [\jcap] {10.1088/1475-7516/2022/04/023}, \href
  {https://ui.adsabs.harvard.edu/abs/2022JCAP...04..023B} {2022, 023}

\bibitem[\protect\citeauthoryear{{Chen} \& {Guestrin}}{{Chen} \&
  {Guestrin}}{2016}]{2016arXiv160302754C}
{Chen} T.,  {Guestrin} C.,  2016, arXiv e-prints, \href
  {https://ui.adsabs.harvard.edu/abs/2016arXiv160302754C} {p. arXiv:1603.02754}

\bibitem[\protect\citeauthoryear{{Chiaro}, {Salvetti}, {La Mura}, {Giroletti},
  {Thompson}  \& {Bastieri}}{{Chiaro} et~al.}{2016}]{2016MNRAS.462.3180C}
{Chiaro} G.,  {Salvetti} D.,  {La Mura} G.,  {Giroletti} M.,  {Thompson} D.~J.,
    {Bastieri} D.,  2016, \mn@doi [\mnras] {10.1093/mnras/stw1830}, \href
  {https://ui.adsabs.harvard.edu/abs/2016MNRAS.462.3180C} {462, 3180}

\bibitem[\protect\citeauthoryear{{Chiaro}, {Kovacevic}  \& {La Mura}}{{Chiaro}
  et~al.}{2021}]{2021JHEAp..29...40C}
{Chiaro} G.,  {Kovacevic} M.,   {La Mura} G.,  2021, \mn@doi [Journal of High
  Energy Astrophysics] {10.1016/j.jheap.2020.11.002}, \href
  {https://ui.adsabs.harvard.edu/abs/2021JHEAp..29...40C} {29, 40}

\bibitem[\protect\citeauthoryear{{Coronado-Bl{\'a}zquez}}{{Coronado-Bl{\'a}zquez}}{2022}]{2022MNRAS.515.1807C}
{Coronado-Bl{\'a}zquez} J.,  2022, \mn@doi [\mnras] {10.1093/mnras/stac1950},
  \href {https://ui.adsabs.harvard.edu/abs/2022MNRAS.515.1807C} {515, 1807}

\bibitem[\protect\citeauthoryear{Dieleman, Willett  \& Dambre}{Dieleman
  et~al.}{2015}]{10.1093/mnras/stv632}
Dieleman S.,  Willett K.~W.,   Dambre J.,  2015, \mn@doi [Monthly Notices of
  the Royal Astronomical Society] {10.1093/mnras/stv632}, 450, 1441

\bibitem[\protect\citeauthoryear{{Finke}, {Kr{\"a}mer}  \& {Manconi}}{{Finke}
  et~al.}{2021}]{2021MNRAS.507.4061F}
{Finke} T.,  {Kr{\"a}mer} M.,   {Manconi} S.,  2021, \mn@doi [\mnras]
  {10.1093/mnras/stab2389}, \href
  {https://ui.adsabs.harvard.edu/abs/2021MNRAS.507.4061F} {507, 4061}

\bibitem[\protect\citeauthoryear{{Fraga}, {Barres de Almeida}, {Bom}, {Brandt},
  {Giommi}, {Schubert}  \& {de Albuquerque}}{{Fraga}
  et~al.}{2021}]{2021MNRAS.505.1268F}
{Fraga} B. M.~O.,  {Barres de Almeida} U.,  {Bom} C.~R.,  {Brandt} C.~H.,
  {Giommi} P.,  {Schubert} P.,   {de Albuquerque} M.~P.,  2021, \mn@doi
  [\mnras] {10.1093/mnras/stab1349}, \href
  {https://ui.adsabs.harvard.edu/abs/2021MNRAS.505.1268F} {505, 1268}

\bibitem[\protect\citeauthoryear{{Germani}, {Tosti}, {Lubrano}, {Cutini},
  {Mereu}  \& {Berretta}}{{Germani} et~al.}{2021}]{2021MNRAS.505.5853G}
{Germani} S.,  {Tosti} G.,  {Lubrano} P.,  {Cutini} S.,  {Mereu} I.,
  {Berretta} A.,  2021, \mn@doi [\mnras] {10.1093/mnras/stab1748}, \href
  {https://ui.adsabs.harvard.edu/abs/2021MNRAS.505.5853G} {505, 5853}

\bibitem[\protect\citeauthoryear{{Ghisellini}, {Maraschi}  \&
  {Treves}}{{Ghisellini} et~al.}{1985}]{1985A&A...146..204G}
{Ghisellini} G.,  {Maraschi} L.,   {Treves} A.,  1985, \aap, \href
  {https://ui.adsabs.harvard.edu/abs/1985A&A...146..204G} {146, 204}

\bibitem[\protect\citeauthoryear{{Glauch}, {Kerscher}  \& {Giommi}}{{Glauch}
  et~al.}{2022}]{2022arXiv220703813G}
{Glauch} T.,  {Kerscher} T.,   {Giommi} P.,  2022, arXiv e-prints, \href
  {https://ui.adsabs.harvard.edu/abs/2022arXiv220703813G} {p. arXiv:2207.03813}

\bibitem[\protect\citeauthoryear{{Golob}, {Sawicki}, {Goulding}  \&
  {Coupon}}{{Golob} et~al.}{2021}]{2021MNRAS.503.4136G}
{Golob} A.,  {Sawicki} M.,  {Goulding} A.~D.,   {Coupon} J.,  2021, \mn@doi
  [\mnras] {10.1093/mnras/stab719}, \href
  {https://ui.adsabs.harvard.edu/abs/2021MNRAS.503.4136G} {503, 4136}

\bibitem[\protect\citeauthoryear{He, Li, Feng, Ho, Ravanbakhsh, Chen  \&
  Póczos}{He et~al.}{2019}]{pnas.1821458116}
He S.,  Li Y.,  Feng Y.,  Ho S.,  Ravanbakhsh S.,  Chen W.,   Póczos B.,
  2019, \mn@doi [Proceedings of the National Academy of Sciences]
  {10.1073/pnas.1821458116}, 116, 13825

\bibitem[\protect\citeauthoryear{{IceCube Collaboration} et~al.,}{{IceCube
  Collaboration} et~al.}{2018a}]{2018Sci...361..147I}
{IceCube Collaboration} et~al., 2018a, \mn@doi [Science]
  {10.1126/science.aat2890}, \href
  {https://ui.adsabs.harvard.edu/abs/2018Sci...361..147I} {361, 147}

\bibitem[\protect\citeauthoryear{{IceCube Collaboration} et~al.,}{{IceCube
  Collaboration} et~al.}{2018b}]{2018Sci...361.1378I}
{IceCube Collaboration} et~al., 2018b, \mn@doi [Science]
  {10.1126/science.aat1378}, \href
  {https://ui.adsabs.harvard.edu/abs/2018Sci...361.1378I} {361, eaat1378}

\bibitem[\protect\citeauthoryear{{Jin}, {Zhang}, {Zhang}, {Zhao}, {Wu}  \&
  {Fan}}{{Jin} et~al.}{2019}]{2019MNRAS.485.4539J}
{Jin} X.,  {Zhang} Y.,  {Zhang} J.,  {Zhao} Y.,  {Wu} X.-b.,   {Fan} D.,  2019,
  \mn@doi [\mnras] {10.1093/mnras/stz680}, \href
  {https://ui.adsabs.harvard.edu/abs/2019MNRAS.485.4539J} {485, 4539}

\bibitem[\protect\citeauthoryear{Ke, Meng, Finley, Wang, Chen, Ma, Ye  \&
  Liu}{Ke et~al.}{2017}]{NIPS2017_6449f44a}
Ke G.,  Meng Q.,  Finley T.,  Wang T.,  Chen W.,  Ma W.,  Ye Q.,   Liu T.-Y.,
  2017, in NIPS.

\bibitem[\protect\citeauthoryear{{Kova{\v{c}}evi{\'c}}, {}, {Chiaro}, {Cutini}
  \& {Tosti}}{{Kova{\v{c}}evi{\'c}} et~al.}{2019}]{2019MNRAS.490.4770K}
{Kova{\v{c}}evi{\'c}} {} M.,  {Chiaro} G.,  {Cutini} S.,   {Tosti} G.,  2019,
  \mn@doi [\mnras] {10.1093/mnras/stz2920}, \href
  {https://ui.adsabs.harvard.edu/abs/2019MNRAS.490.4770K} {490, 4770}

\bibitem[\protect\citeauthoryear{{Kova{\v{c}}evi{\'c}}, {Chiaro}, {Cutini}  \&
  {Tosti}}{{Kova{\v{c}}evi{\'c}} et~al.}{2020}]{2020MNRAS.493.1926K}
{Kova{\v{c}}evi{\'c}} M.,  {Chiaro} G.,  {Cutini} S.,   {Tosti} G.,  2020,
  \mn@doi [\mnras] {10.1093/mnras/staa394}, \href
  {https://ui.adsabs.harvard.edu/abs/2020MNRAS.493.1926K} {493, 1926}

\bibitem[\protect\citeauthoryear{{Lefaucheur} \& {Pita}}{{Lefaucheur} \&
  {Pita}}{2017}]{2017A&A...602A..86L}
{Lefaucheur} J.,  {Pita} S.,  2017, \mn@doi [\aap]
  {10.1051/0004-6361/201629552}, \href
  {https://ui.adsabs.harvard.edu/abs/2017A&A...602A..86L} {602, A86}

\bibitem[\protect\citeauthoryear{{Narendra}, {Gibson}, {Dainotti}, {Bogdan},
  {Pollo}, {Liodakis}, {Poliszczuk}  \& {Rinaldi}}{{Narendra}
  et~al.}{2022}]{2022ApJS..259...55N}
{Narendra} A.,  {Gibson} S.~J.,  {Dainotti} M.~G.,  {Bogdan} M.,  {Pollo} A.,
  {Liodakis} I.,  {Poliszczuk} A.,   {Rinaldi} E.,  2022, \mn@doi [\apjs]
  {10.3847/1538-4365/ac545a}, \href
  {https://ui.adsabs.harvard.edu/abs/2022ApJS..259...55N} {259, 55}

\bibitem[\protect\citeauthoryear{{Padovani}, {Giommi}, {Resconi}, {Glauch},
  {Arsioli}, {Sahakyan}  \& {Huber}}{{Padovani}
  et~al.}{2018}]{2018MNRAS.480..192P}
{Padovani} P.,  {Giommi} P.,  {Resconi} E.,  {Glauch} T.,  {Arsioli} B.,
  {Sahakyan} N.,   {Huber} M.,  2018, \mn@doi [\mnras] {10.1093/mnras/sty1852},
  \href {https://ui.adsabs.harvard.edu/abs/2018MNRAS.480..192P} {480, 192}

\bibitem[\protect\citeauthoryear{{Sahakyan}, {Giommi}, {Padovani},
  {Petropoulou}, {B{\'e}gu{\'e}}, {Boccardi}  \& {Gasparyan}}{{Sahakyan}
  et~al.}{2022}]{2022arXiv220405060S}
{Sahakyan} N.,  {Giommi} P.,  {Padovani} P.,  {Petropoulou} M.,
  {B{\'e}gu{\'e}} D.,  {Boccardi} B.,   {Gasparyan} S.,  2022, arXiv e-prints,
  \href {https://ui.adsabs.harvard.edu/abs/2022arXiv220405060S} {p.
  arXiv:2204.05060}

\bibitem[\protect\citeauthoryear{{Salvetti}, {Chiaro}, {La Mura}  \&
  {Thompson}}{{Salvetti} et~al.}{2017}]{2017MNRAS.470.1291S}
{Salvetti} D.,  {Chiaro} G.,  {La Mura} G.,   {Thompson} D.~J.,  2017, \mn@doi
  [\mnras] {10.1093/mnras/stx1328}, \href
  {https://ui.adsabs.harvard.edu/abs/2017MNRAS.470.1291S} {470, 1291}

\bibitem[\protect\citeauthoryear{{Saz Parkinson}, {Xu}, {Yu}, {Salvetti},
  {Marelli}  \& {Falcone}}{{Saz Parkinson} et~al.}{2016}]{2016ApJ...820....8S}
{Saz Parkinson} P.~M.,  {Xu} H.,  {Yu} P.~L.~H.,  {Salvetti} D.,  {Marelli} M.,
    {Falcone} A.~D.,  2016, \mn@doi [\apj] {10.3847/0004-637X/820/1/8}, \href
  {https://ui.adsabs.harvard.edu/abs/2016ApJ...820....8S} {820, 8}

\bibitem[\protect\citeauthoryear{{Urry} \& {Padovani}}{{Urry} \&
  {Padovani}}{1995}]{1995PASP..107..803U}
{Urry} C.~M.,  {Padovani} P.,  1995, \mn@doi [\pasp] {10.1086/133630}, \href
  {https://ui.adsabs.harvard.edu/abs/1995PASP..107..803U} {107, 803}

\bibitem[\protect\citeauthoryear{{Xu}, {Huang}, {Deng}, {Mei}  \& {Wang}}{{Xu}
  et~al.}{2020}]{2020ApJ...895..133X}
{Xu} Y.,  {Huang} W.,  {Deng} H.,  {Mei} Y.,   {Wang} F.,  2020, \mn@doi [\apj]
  {10.3847/1538-4357/ab8ae3}, \href
  {https://ui.adsabs.harvard.edu/abs/2020ApJ...895..133X} {895, 133}

\bibitem[\protect\citeauthoryear{{Yi}, {Chen}, {Pan}, {Yue}, {Lu}, {Li}  \&
  {Luo}}{{Yi} et~al.}{2019}]{2019ApJ...887..241Y}
{Yi} Z.,  {Chen} Z.,  {Pan} J.,  {Yue} L.,  {Lu} Y.,  {Li} J.,   {Luo} A.~L.,
  2019, \mn@doi [\apj] {10.3847/1538-4357/ab54d0}, \href
  {https://ui.adsabs.harvard.edu/abs/2019ApJ...887..241Y} {887, 241}

\bibitem[\protect\citeauthoryear{{Zhu}, {Kang}  \& {Zheng}}{{Zhu}
  et~al.}{2021}]{2021RAA....21...15Z}
{Zhu} K.-R.,  {Kang} S.-J.,   {Zheng} Y.-G.,  2021, \mn@doi [Research in
  Astronomy and Astrophysics] {10.1088/1674-4527/21/1/15}, \href
  {https://ui.adsabs.harvard.edu/abs/2021RAA....21...15Z} {21, 015}

\makeatother
\end{thebibliography}


 





\bsp	
\label{lastpage}
\end{document}